\documentclass[a4paper,fleqn]{cas-dc}

\usepackage[numbers]{natbib}

\usepackage{amsmath}
\usepackage{amsthm}
\usepackage{amssymb}
\usepackage{xcolor}
\usepackage{subcaption}

\newtheorem{mdef}{Definition}
\newtheorem{prop}{Proposition}

\DeclareMathOperator{\sig}{sig}
\DeclareMathOperator*{\argmin}{arg\,min}

\DeclareMathOperator{\sd}{sd}
\DeclareMathOperator{\MSE}{MSE}

\usepackage{tikz}
\usepackage{pgfplots}
\usepgfplotslibrary{statistics}

\graphicspath{{largeScale/}{img/error-LPS-Plots/}}

\definecolor{db}{rgb}{0.035, 0.239, 0.498} 
\definecolor{b}{rgb}{0.211, 0.45, 0.76} 
\definecolor{lb}{rgb}{0.447, 0.69, 1.} 
\definecolor{o}{rgb}{0.8, 0.588, 0.204} 

\newcommand{\rev}[1]{\textcolor{black}{#1}}

\newcounter{mparcnt}

\newcommand\mpar[1]{}

\usepackage{marginnote}

\begin{document}

\let\WriteBookmarks\relax
\def\floatpagepagefraction{1}
\def\textpagefraction{.001}
\shorttitle{Evaluation of Neighborhood Weights and Sizes}
\shortauthors{M. Skrodzki and E. Zimmermann}

\title [mode = title]{A Large-Scale Evaluation of Shape-Aware Neighborhood Weights and Neighborhood Sizes}
\tnotemark[1]


\tnotetext[1]{This material is based upon work supported by the National Science Foundation under Grant No. DMS-1439786 and the Alfred P. Sloan Foundation award G-2019-11406 while the author was in residence at the Institute for Computational and Experimental Research in Mathematics in Providence, RI, during the Illustrating Mathematics program.
This research was partially funded by the Deutsche Forschungsgemeinschaft (DFG, German Research Foun-
dation) under grant number 455095046 and as part of the collaborative research cluster SFB Transregio 109, ``Discretization in Geometry and Dynamics''.
Finally, this work was supported by the German National Academic Foundation.}

\author[1]{Martin Skrodzki}[
						bioid=1,
                        orcid=0000-0002-8126-0511,
                        twitter=msmathcomputer2]
\cormark[1]
\ead{mail@ms-math-computer.science}
\ead[url]{https://ms-math-computer.science}

\credit{Conceptualization of this study, Methodology, Software, Evaluation}

\address[1]{ICERM, Brown University, Providence, RI, USA and RIKEN iTHEMS, Wako, Saitama, Japan and CGV, TU Delft, Delft, the Netherlands}

\author[2]{Eric Zimmermann}[
						bioid=2]
\ead{eric.zimmermann@fu-berlin.de}
\ead[URL]{https://userpage.fu-berlin.de/ezimmermann/}

\credit{Conceptualization of this study, Methodology, Software, Evaluation}

\address[2]{Institute of Mathematics, Freie Universit\"at Berlin, Berlin, Germany}

\cortext[cor1]{Corresponding author}

\begin{abstract}
In this paper, we define and evaluate a weighting scheme for neighborhoods in point sets.
Our weighting takes the shape of the geometry, i.e., the normal information, into account.
This causes the obtained neighborhoods to be more reliable in the sense that connectivity also depends on the orientation of the point set. 
We utilize a sigmoid to define the weights based on the normal variation. 
For an evaluation of the weighting scheme, we turn to a Shannon entropy model for feature \rev{classification that can be proven to be non-degenerate} for our family of weights. 
Based on this model, we evaluate our weighting terms on a large scale of both clean and real-world models.
This evaluation provides results regarding the choice of optimal parameters within our weighting scheme.
Furthermore, the large-scale evaluation also reveals that neighborhood sizes should not be fixed globally when processing models.
\rev{Finally, we highlight the applicability of our weighting scheme withing the application context of denoising.}
\end{abstract}

\begin{graphicalabstract}
	\includegraphics[width=0.28\textwidth]{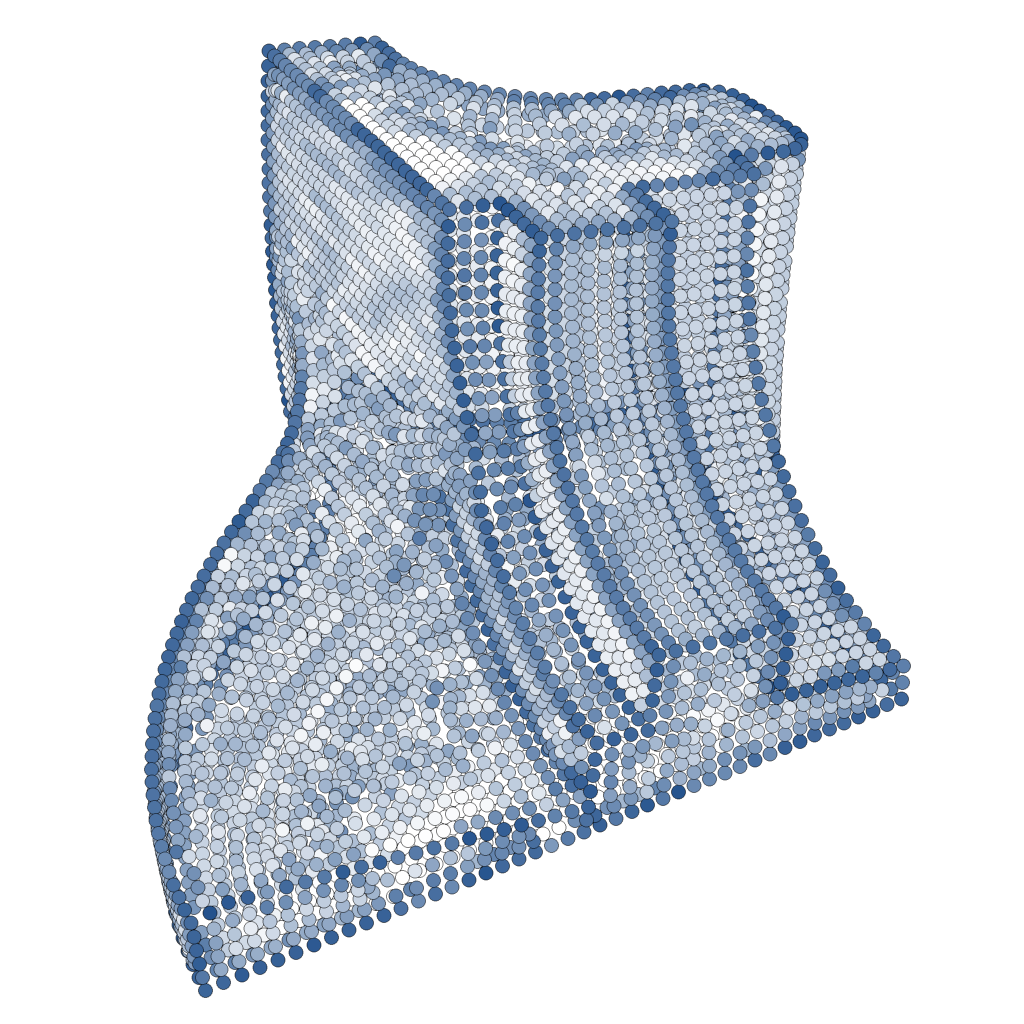}
	\hfill
	\includegraphics[width=0.28\textwidth]{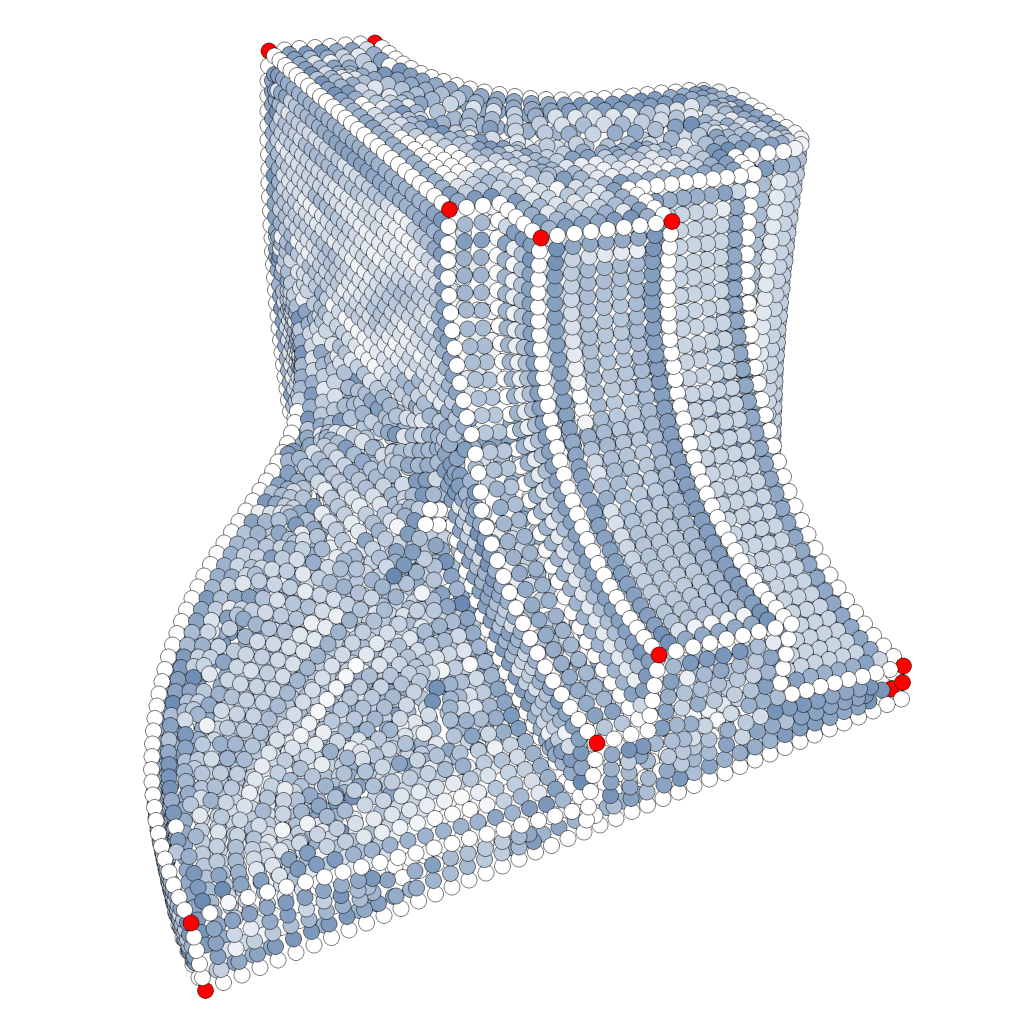}
	\hfill
	\includegraphics[width=0.28\textwidth]{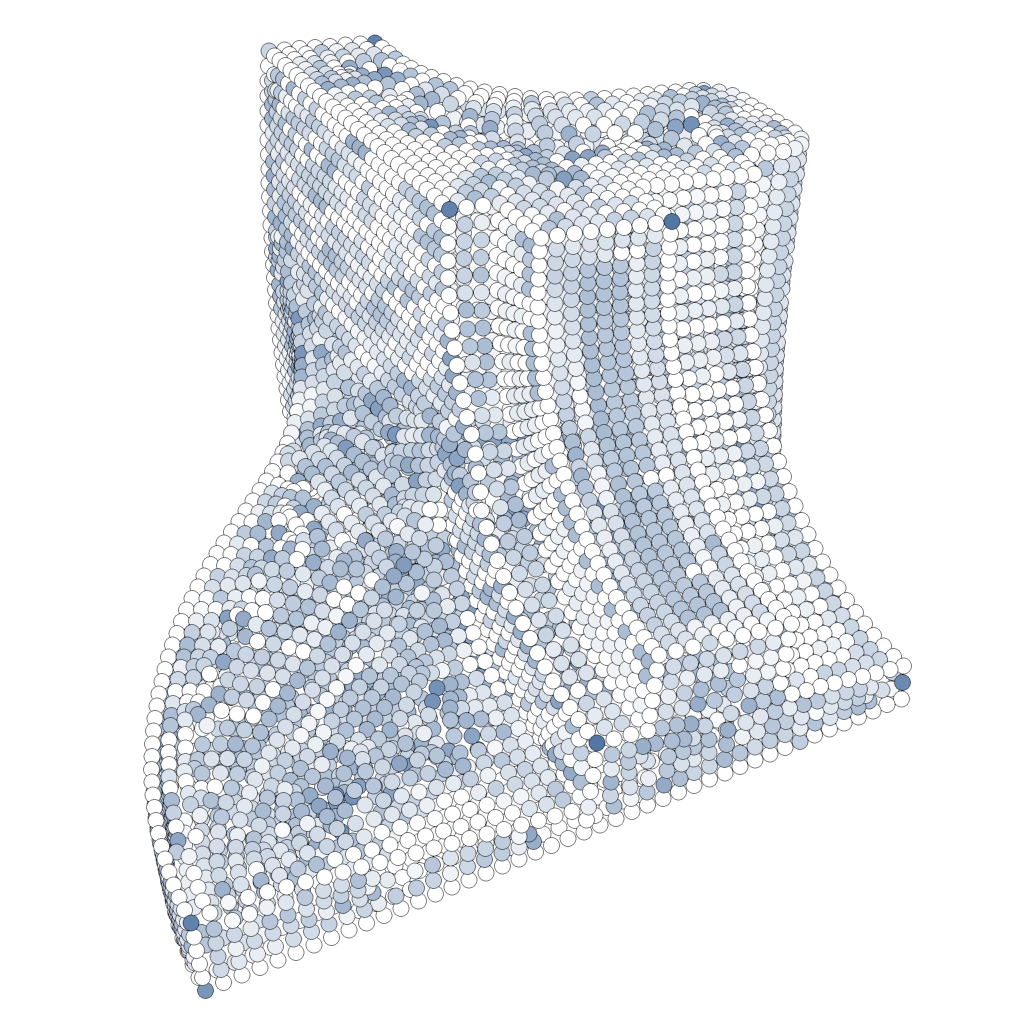}
	\hfill
	\def\svgwidth{0.07\textwidth}
\begingroup%
  \makeatletter%
  \providecommand\color[2][]{%
    \errmessage{(Inkscape) Color is used for the text in Inkscape, but the package 'color.sty' is not loaded}%
    \renewcommand\color[2][]{}%
  }%
  \providecommand\transparent[1]{%
    \errmessage{(Inkscape) Transparency is used (non-zero) for the text in Inkscape, but the package 'transparent.sty' is not loaded}%
    \renewcommand\transparent[1]{}%
  }%
  \providecommand\rotatebox[2]{#2}%
  \newcommand*\fsize{\dimexpr\f@size pt\relax}%
  \newcommand*\lineheight[1]{\fontsize{\fsize}{#1\fsize}\selectfont}%
  \ifx\svgwidth\undefined%
    \setlength{\unitlength}{212.10440808bp}%
    \ifx\svgscale\undefined%
      \relax%
    \else%
      \setlength{\unitlength}{\unitlength * \real{\svgscale}}%
    \fi%
  \else%
    \setlength{\unitlength}{\svgwidth}%
  \fi%
  \global\let\svgwidth\undefined%
  \global\let\svgscale\undefined%
  \makeatother%
  \begin{picture}(1,3.66008946)%
    \lineheight{1}%
    \setlength\tabcolsep{0pt}%
    \put(0,0){\includegraphics[width=\unitlength,page=1]{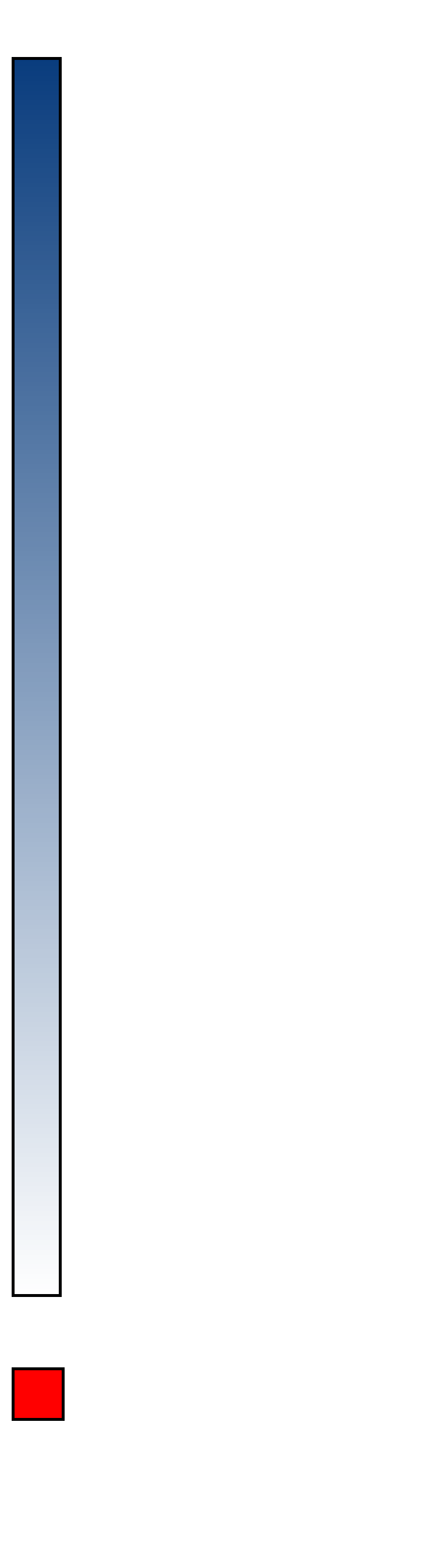}}%
    \put(0.22589176,3.4173939){\color[rgb]{0,0,0}\makebox(0,0)[lt]{\lineheight{1.25}\smash{\begin{tabular}[t]{l}$3e^{-1}$\end{tabular}}}}%
    \put(0.21809134,0.65331303){\color[rgb]{0,0,0}\makebox(0,0)[lt]{\lineheight{1.25}\smash{\begin{tabular}[t]{l}$0$\end{tabular}}}}%
    \put(0.15869317,2.04636901){\color[rgb]{0,0,0}\makebox(0,0)[lt]{\lineheight{1.25}\smash{\begin{tabular}[t]{l}$E_i^{\dim}$\end{tabular}}}}%
    \put(0.19030206,0.32737375){\color[rgb]{0,0,0}\makebox(0,0)[lt]{\lineheight{1.25}\smash{\begin{tabular}[t]{l}$C_i\text{ fail}$\end{tabular}}}}%
  \end{picture}%
\endgroup%

	\\
	\includegraphics[width=0.28\textwidth]{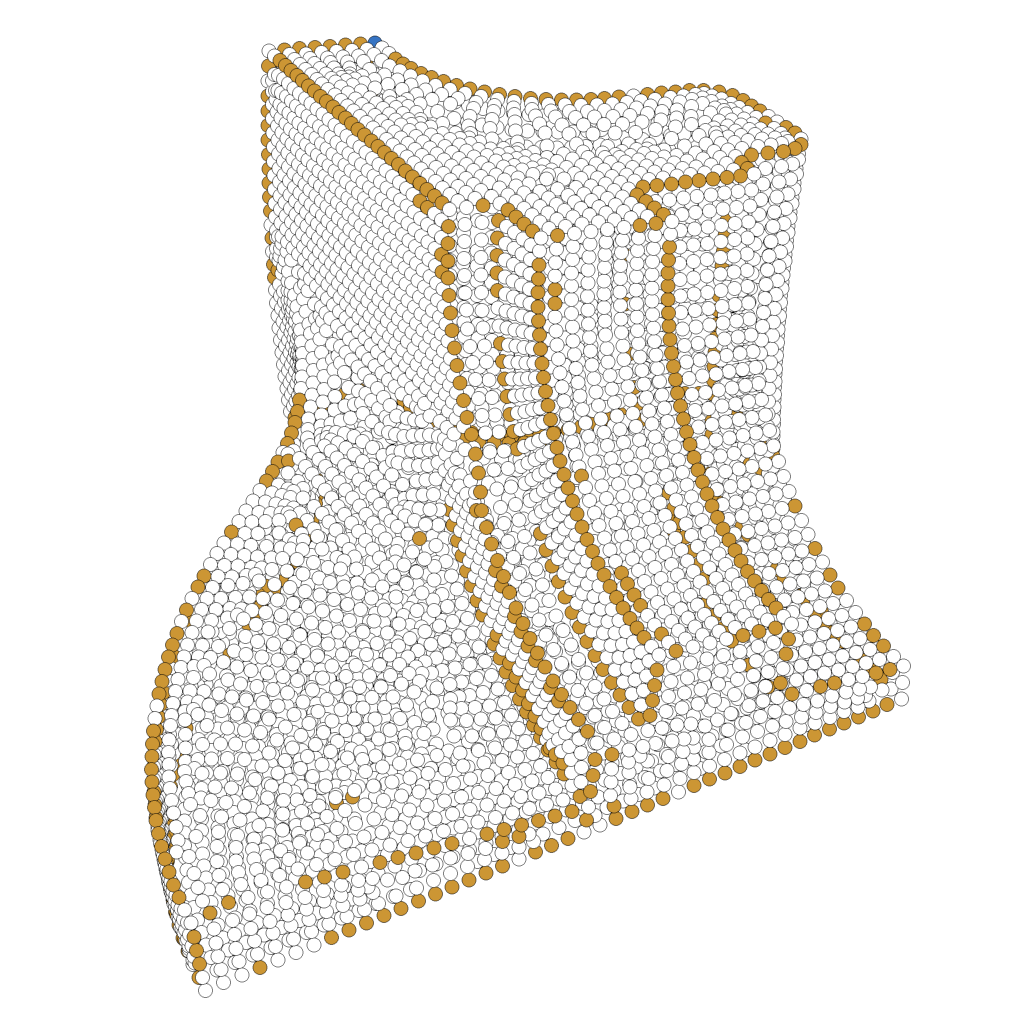}
	\hfill
	\includegraphics[width=0.28\textwidth]{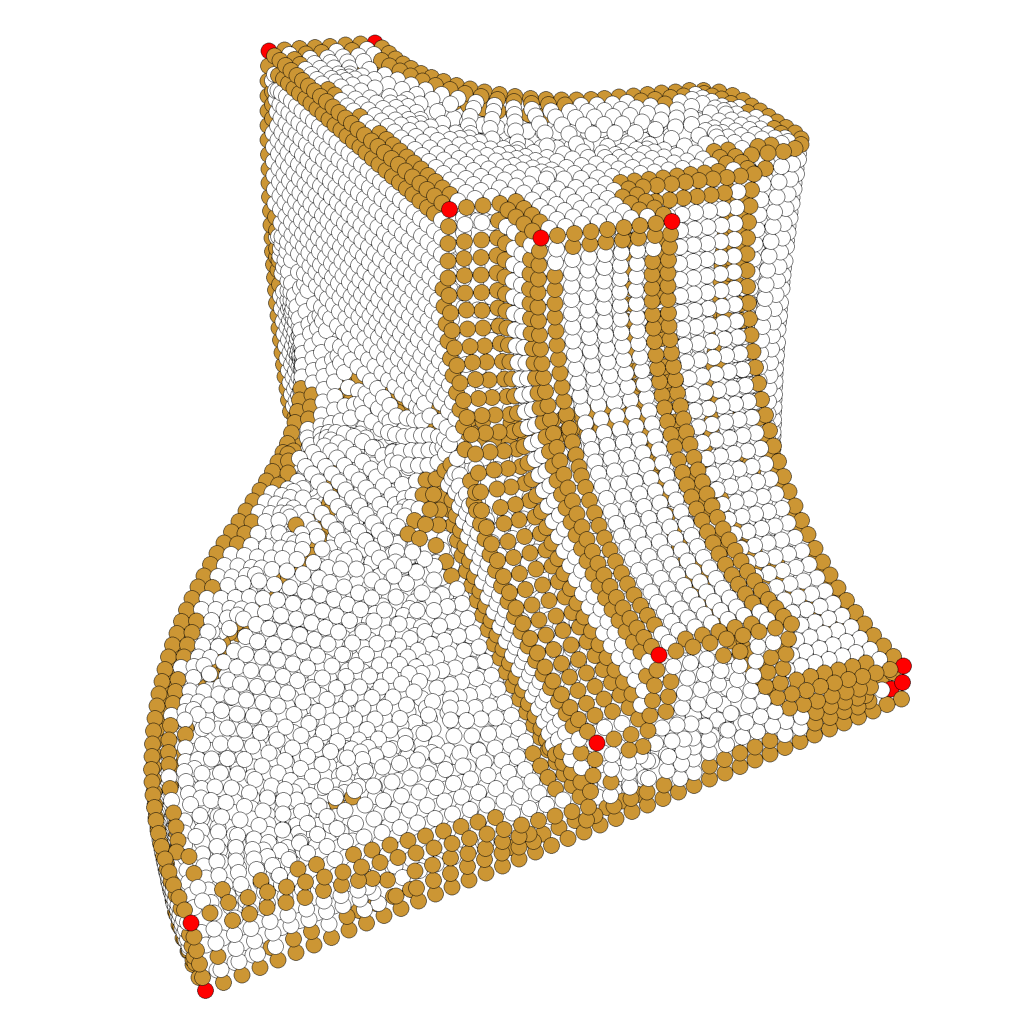}
	\hfill
	\includegraphics[width=0.28\textwidth]{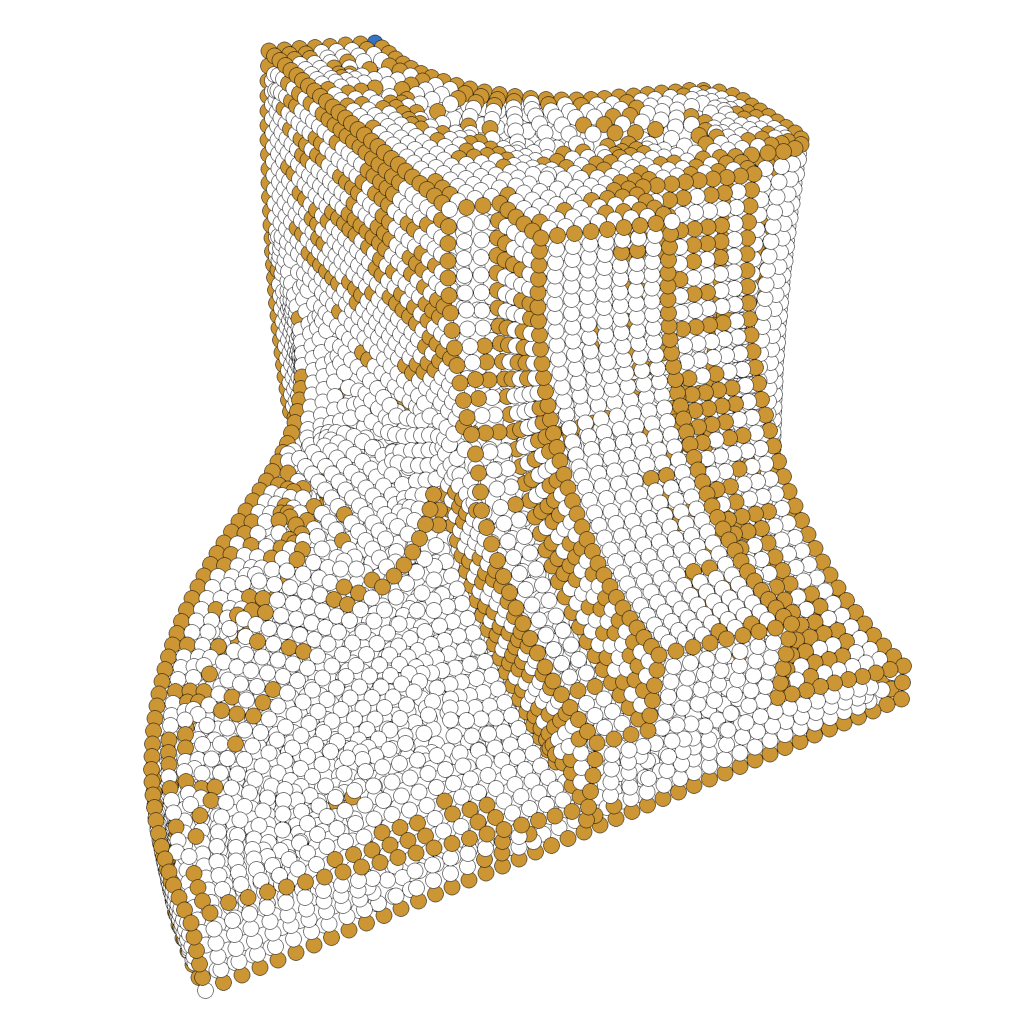}
	\def\svgwidth{0.07\textwidth}
\begingroup%
  \makeatletter%
  \providecommand\color[2][]{%
    \errmessage{(Inkscape) Color is used for the text in Inkscape, but the package 'color.sty' is not loaded}%
    \renewcommand\color[2][]{}%
  }%
  \providecommand\transparent[1]{%
    \errmessage{(Inkscape) Transparency is used (non-zero) for the text in Inkscape, but the package 'transparent.sty' is not loaded}%
    \renewcommand\transparent[1]{}%
  }%
  \providecommand\rotatebox[2]{#2}%
  \newcommand*\fsize{\dimexpr\f@size pt\relax}%
  \newcommand*\lineheight[1]{\fontsize{\fsize}{#1\fsize}\selectfont}%
  \ifx\svgwidth\undefined%
    \setlength{\unitlength}{212.10440808bp}%
    \ifx\svgscale\undefined%
      \relax%
    \else%
      \setlength{\unitlength}{\unitlength * \real{\svgscale}}%
    \fi%
  \else%
    \setlength{\unitlength}{\svgwidth}%
  \fi%
  \global\let\svgwidth\undefined%
  \global\let\svgscale\undefined%
  \makeatother%
  \begin{picture}(1,3.66008946)%
    \lineheight{1}%
    \setlength\tabcolsep{0pt}%
    \put(0,0){\includegraphics[width=\unitlength,page=1]{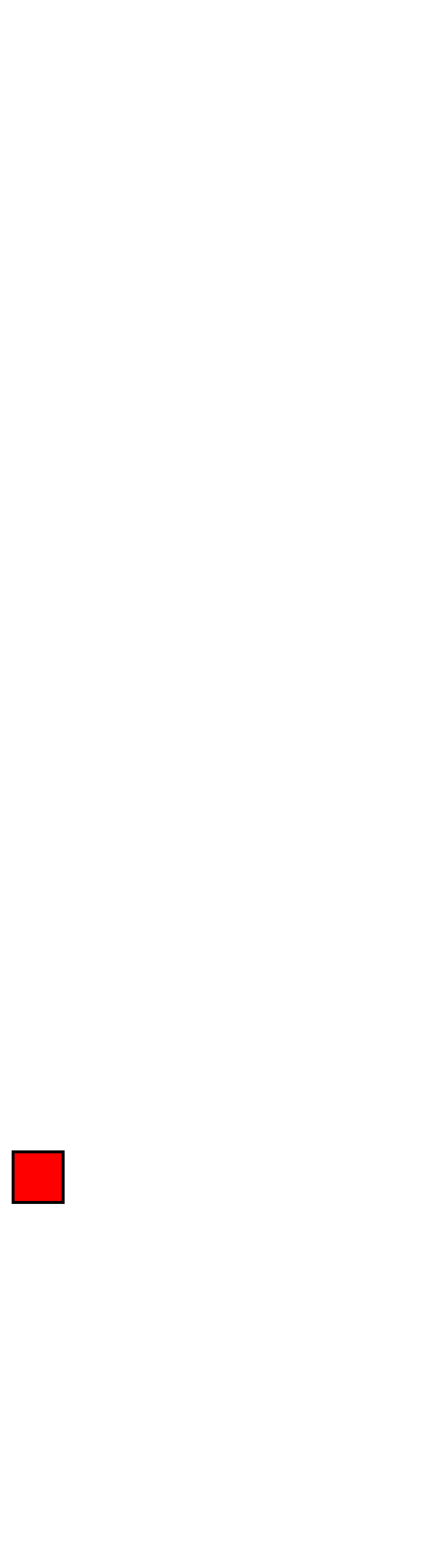}}%
    \put(0.19030206,0.83398901){\color[rgb]{0,0,0}\makebox(0,0)[lt]{\lineheight{1.25}\smash{\begin{tabular}[t]{l}$C_i \text{ fail}$\end{tabular}}}}%
    \put(0,0){\includegraphics[width=\unitlength,page=2]{colorbar2.pdf}}%
    \put(0.18606517,1.57248383){\color[rgb]{0,0,0}\makebox(0,0)[lt]{\lineheight{1.25}\smash{\begin{tabular}[t]{l}$\text{scatter}$\end{tabular}}}}%
    \put(0,0){\includegraphics[width=\unitlength,page=3]{colorbar2.pdf}}%
    \put(0.18606517,2.3223251){\color[rgb]{0,0,0}\makebox(0,0)[lt]{\lineheight{1.25}\smash{\begin{tabular}[t]{l}$\text{planar}$\end{tabular}}}}%
    \put(0,0){\includegraphics[width=\unitlength,page=4]{colorbar2.pdf}}%
    \put(0.18606517,3.07920619){\color[rgb]{0,0,0}\makebox(0,0)[lt]{\lineheight{1.25}\smash{\begin{tabular}[t]{l}$\text{linear}$\end{tabular}}}}%
  \end{picture}%
\endgroup%

\end{graphicalabstract}

\begin{highlights}
\item Definition of a shape-aware neighborhood weighting utilizing sigmoid function weights based on normal variation;
\item Presentation of a \rev{Shannon entropy evaluation model that can be proven to be non-degenerate on our inputs};
\item Large scale experimental evaluation of the proposed neighborhood weighting concept;
\item Discussion of the results with respect to both neighborhood weighting and neighborhood sizes.
\end{highlights}


\maketitle

\section{Introduction}

\noindent Point sets arise naturally in many kinds of 3D acquisition processes, like, e.g., 3D laser-scanning. As early as 1985, they have been recognized as fundamental shape representations in computer graphics, see~\cite{levoy1985use}. Ever since, they have been used in diverse applications, e.g., in archaeology~\cite{levoy2000digital}, face recognition~\cite{boehnen2005accuracy}, or traffic accident analysis~\cite{buck2007application}.

Despite their versatility and their advantages---like easy acquisition and low storage costs---point sets have a significant downside to them when compared with mesh representations: They are not equipped with connectivity information. This is mostly due to the acquisition process. Consider for example a manually guided scanning device. The operator will scan those areas of the real-world objects with very sharp features multiple times. Consequently, occlusion is prevented and the whole geometry is captured. Even though each scan can provide connectivity information on the respectively acquired points, the complete point set obtained via registration of the individual scans (see, e.g.,~\cite{bellekens2014survey}) does not provide global connectivity information in general. Thus, a notion of neighborhoods has to be defined and computed for each point.

Many definitions of neighborhoods, combinatorial or geometric, with global or local parameters, have been proposed and discussed (see Section~\ref{sec:RelatedWork}). Furthermore, the concept of weighting neighboring points is not new. For example, the pure selection of a neighborhood causes an equal treatment of all neighbors. Aside from this, isotropic weighting is one common way, evaluating Euclidean distances via a Gaussian weighting function. This provides closer points with higher influence (see, e.g.,~\cite{alexa2001point}). Additionally, other point set information can be incorporated, like density or distribution (see, e.g.,~\cite{park2012multi} or~\cite{skrodzki2018densityWeights}). The inclusion of normal deviation in the area of anisotropic weighting has also been considered and discussed before (see~\cite{yadav2018constraint,skrodzki2019neighborhood}).

The research work presented here aims at investigating anisotropic weighting terms in a broad framework, which includes usual weighting choices such as equal weights or sharp cut-off weights\footnote{We consider the case of cut-off weights if starting from a given deviation, all points with greater or equal deviation are attributed weight~$0$.} (Section~\ref{sec:Method}).
Our evaluation is processed via a Shannon entropy model (Section \ref{sec:Evaluation}), which is based on the work of~\cite{demantke2011dimensionality,weinmann2014semantic}.
Furthermore, we aim at evaluating the weighting scheme on a large scale. This is to prevent over-interpretation of findings obtained from a very small set of models.
Overall, the contributions of this work are:
\begin{itemize}
	\item Definition of a shape-aware neighborhood weighting utilizing sigmoid function weights based on normal variation;
	\item Presentation of a \rev{Shannon entropy evaluation model that can be proven to be non-degenerate on our inputs};
	\item Large scale experimental evaluation of the proposed neighborhood weighting concept;
	\item Discussion of the results with respect to both neighborhood weighting and neighborhood sizes.
\end{itemize}
\mpar{\label{rev:32}}\rev{While the content of this paper is deeply routed in the field of traditional computer-aided design, in our concluding Section~\ref{sec:conclusion}, we will provide an outlook and several thoughts on the application of the presented techniques within the context of machine learning.}


\section{Related Work}
\label{sec:RelatedWork}

\noindent Neighborhoods are very important in point set processing, as almost all algorithmic approaches rely on them.
A common choice is to use heuristics to determine sufficient notions like the size of a combinatorial or metric neighborhood. 
In the following, we recall works discussing heuristic neighborhood definitions.
Several works have advanced from simple heuristics and derive more involved notions for better fitting neighborhood definitions in different contexts.
These are mainly obtained from error functionals, which we will also discuss.

\subsection{Heuristics}
\label{sec:Heuristics}

\noindent Most works consider either a combinatorial $k$-nearest neighborhood~$\mathcal{N}_k(\cdot)$ or a metric ball~$B_r(\cdot)$ inducing a neighborhood.
Both of these notions have parameters to be tuned, namely the number of neighbors~$k$ or the radius~$r$ of the neighborhood. 
Several works have been presented introducing heuristics to find appropriate values for~$k$ or~$r$ in different scenarios.
The authors of~\cite{alexa2001point} for instance use a global radius and change it to affect the running time of their algorithm. 
In~\cite{pauly2002efficient}, the authors fix a combinatorial number~$k$ of neighbors to be sought.
Then, for each point~$p_i$ from the considered point set~$P$, these~$k$ neighbors are found, which fixes a radius~$r_i$ to the farthest of them. Finally, the neighbors within radius~$r_i/3$ are used. 
Therefore, their approach resembles the geometric neighborhood in a local manner.

The method used in~\cite{pauly2003shape} is more involved. 
The authors recognize that both a too large or too small radius~$r$ lead to problems and thus aim for a local adaption like~\cite{pauly2002efficient}. 
A local density estimate~$\delta_i$ around each point~$ {p_i\in P} $ is computed from the smallest ball centered at~$p_i$, containing~$\mathcal{N}_k(p_i)$, where~$k$ is found experimentally to be best chosen from~\mbox{$\{6,\ldots,20\}\subset\mathbb{N}$}. 
Given the radius~$r_i$ of this ball, the local density is set to be~$ {\delta_i=k/r_i^2} $. 
In a second step, a smooth density function~$\delta$ is interpolated from the local density estimates~$\delta_i$, hence this weighting involves the incorporation of density-information into the weight assignment.

In the context of surface reconstruction, the authors of~\cite{floater2001meshless} discuss several choices for neighborhoods and corresponding weights. 
While two of the three presented methods simply use geometric neighborhoods, the third method takes a different approach. 
Namely, the authors collect all neighbors of~$ p_i $ in a ``large'' ball~(\cite[page~7]{floater2001meshless}) around~$p_i$. Then, they fit a plane to this preliminary neighborhood and project all neighbors and~$p_i$ onto this plane. On the projections, a Delaunay triangulation is built and the induced neighborhood of the triangulation is used in the following computations, which localizes their approach and respects different point distributions.

A completely different route is taken by~\cite{brodu2012terrestrial}. The authors first calculate features of a point set based on differently sized neighborhoods. Then, they use a training procedure to find the combination of neighborhood sizes that provides the best separation of different feature classes.

The inclusion of normal deviation and hence anisotropic weighting into neighborhood concepts is part of the work~\cite{yadav2018constraint}. 
The approach of the authors is to use a weighted principal component analysis, which fits our evaluation model.
However, they rely on a global neighborhood size and assign sharp cut-off weights while we allow for changing neighborhood sizes and smooth weighting terms.

\subsection{Error Functionals}
\label{sec:OptimalneighborhoodSizesFromErrorFunctionals}

\noindent While the approaches presented above are based on heuristics, some works try to deduce an optimal~$k$ for the~$k$ nearest neighborhoods based on error functions. 
For instance, the authors of~\cite{lipman2006error} work in the context of the \mpar{\label{rev:35}}Moving Least Squares (MLS) framework (see~\cite{alexa2001point,levin1998approximation,levin2004mesh,sober2016manifolds}) for function approximation. 
The authors perform an extensive error analysis to quantify the approximation error both independent and depending on the given data. 
Finally, they obtain an error functional. 
This is then evaluated for different neighborhood sizes~$k$. 
The neighborhood~$\mathcal{N}_k$ yielding the smallest error is finally chosen to be used in the actual MLS approximation.

In contrast, the authors of~\cite{mitra2004estimating} deduce an error bound on the normal estimation obtained from different neighborhood sizes. 
Utilizing this error functional, they obtain the best suited neighborhood size for normal computation.
The work of~\cite{lipman2006error} heavily depends on the MLS framework in which the error analysis is deduced, while the work of~\cite{mitra2004estimating} depends on the framework of normal computation. 

The authors of~\cite{weinmann2014semantic} take a more general approach in the context of segmentation of 3D point sets. They also use the concept of combinatorial neighborhoods, going back to results of~\cite{linsen2001local,demantke2011dimensionality}. 
In order to choose an optimal value for~$k$, the authors turn to the covariance matrix, which is symmetric and positive-semi-definite. 
Thus, the matrix has three non-negative eigenvalues.
Following an idea of~\cite{hoppe1992surface}, in the work of~\cite{pauly2003shape}, the authors grow a neighborhood and consider a surface variation as a measure to grow a neighborhood around each point~$p_i$. 
The same quantity is used by~\cite{belton2006classification}. 
However, the authors of~\cite{pauly2003shape} do not grow a neighborhood, but choose a size~$k$ for it according to a consistent curvature level.
The authors of~\cite{weinmann2014semantic} do not stop at these information, but proceed to consider three more quantities derived from the eigenvalues of the covariance matrix reflecting point set features, see~\cite{demantke2011dimensionality,weinmann2014semantic}. 
Afterwards, following the concept of entropy by Shannon~\cite{shannon1948mathematical}, they evaluate combinatorial and geometric neighborhood sizes via two error measures (see Section~\ref{sec:Evaluation} for a detailed discussion).


\section{Sigmoid Weights}
\label{sec:Method}

\noindent\mpar{\label{rev:36}}\rev{In contrast to the works listed above, our approach aims at integrating the shape of the geometry, i.e., the normal information, into the neighborhood definition.
We will do so by enriching a given combinatorial neighborhood with a set of weights that are dependent on the normal variation within the neighborhood.
To ensure a smooth transition of weights, we apply a sigmoid function to the angle deviation of the normals.}

\noindent Given a set of points~$ {P = \lbrace p_i \mid i\in[n] \rbrace} $, $n\in\mathbb{N}$, corresponding oriented unit-length normals~$ {n_i\in\mathbb{S}^2} $, and local neighborhoods~$ {\mathcal{N}_i\subset [n]} $ for every~$ {i \in [n]} $. 
For a given weighting function
\begin{equation}
	\phi : [0,\rev{\pi}] \to [0,1],
\end{equation}
we obtain the following weights
\begin{eqnarray}
\label{equ:Weights}
w_{ij} = \phi \left(\rev{\angle(n_i,n_j)}\right) \quad \text{ for } i \in [n],\ j \in \mathcal{N}_i.
\end{eqnarray}
\rev{The argument of~$ \phi $ is the deviation of the normals measured by their angle, which ranges from~$ 0 $ to~$ \pi $.
We turn to this formulation, because it has an obvious geometric interpretation.
In order to have an efficient implementation of the presented techniques, the scenario can be reformulated in terms of the scalar product of the normals, which avoids the costly computation of~$ \arccos $.
}

Note that by the symmetry of the \rev{angle,} the weights are symmetric, i.e.,~$ {w_{ij} = w_{ji}} $. 
The weighting function~$ \phi $ shall assign non-negative weights between~$ 0 $ and~$ 1 $. 
These weights should correspond to the similarity of the corresponding normals, i.e., a small \rev{angle} should result in weights close to \rev{or equal}~$ 1 $, while a \rev{large angle} should yield weights close to \rev{or equal}~$ 0 $.

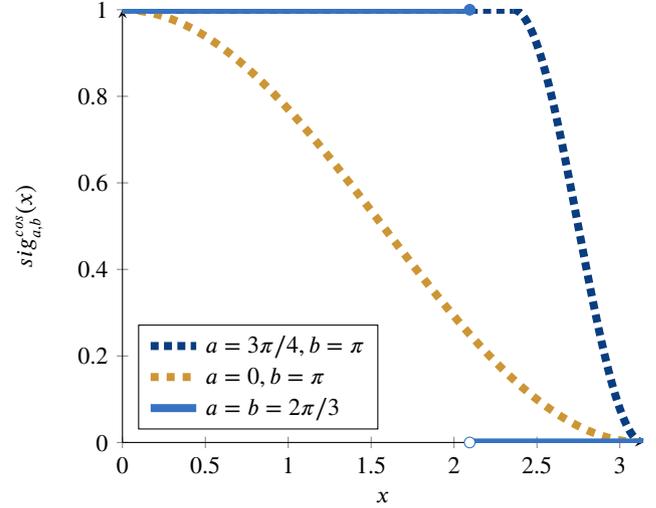
\begin{figure}
	\centering
	\begin{tikzpicture}
	\pgfplotsset{		
		samples=150,
		axis lines=left,
	}
	\begin{axis}[
	legend pos = south west,
	xlabel={$x$},
	ylabel={$sig^{cos}_{a,b} (x)$},
	legend cell align = {left},
	]
	\addplot[db, line width=3pt, dotted, domain=0:3*pi/4] {1};
	\addplot[db, line width=3pt, dotted, domain=3*pi/4:pi] {0.5*cos(deg(pi*(\x-(3*pi/4)) / (pi-(3*pi/4)))) + 0.5};
	
	\addplot[o, line width=3pt, dashed, domain=0:pi] {0.5*cos(deg(pi*(\x-0.) / (pi))) + 0.5};

	\addplot[b, line width=3pt, domain=0:2*pi/3] {1};
	\addplot[b, mark=*,fill=white,only marks] coordinates {(2*pi/3,0)};
	\addplot[b, mark=*,only marks] coordinates {(2*pi/3,1)};
	\addplot[b, line width=3pt, domain=2*pi/3:pi] {0};
	
	\legend{{}{$ a = 3\pi/4, b = \pi $},,{}{$ a = 0, b = \pi $},,,,{}{$ a = b = 2\pi/3 $}}
	
	\end{axis}
	\end{tikzpicture}
	\caption{Plots of the sigmoid $ sig^{cos}_{a,b} (x)$ for three parameter choices \rev{on the domain~$[0,\pi]$}.}
	\label{fig:sigcosExample}
\end{figure}

Our choice for the weighting function is a sigmoid.
A sigmoid function is visually characterized by its shape of an ``S''-curve, \rev{even though it is mirrored in our scenario,} see Figure~\ref{fig:sigcosExample}.
We will consider a family of sigmoid functions that provide different interpolations between~$ 1 $ and~$ 0 $.
The family is based on the trigonometric cosine function.
It is related to the sigmoid used in~\cite{marler2006sigmoidally}, however, \rev{we alter it to be a monotonic falling curve between~$ 1 $ and~$ 0 $ on the interval~$ [0,\pi] $}.
\begin{mdef}[Cosine-Sigmoid]
	\label{def:sigmoid_cos}
	\mpar{\label{rev:17}} Consider two given thresholds $ \rev{a \in \left[0, \pi\right]} $ and $ \rev{b \in \left[a, \pi\right]} $. Then, we define the sigmoid weighting function~$ \sig^{\cos}_{a,b}:[0,\rev{\pi}]\rightarrow[0,1] $ as
	\mpar{\label{rev:21}}
	\begin{align}
		\label{equ:sigmoid_cos}
		x \mapsto 
		\rev{\begin{cases} 
			1 & x \in \left[0, a\right[,\\ \frac{1}{2}\cos\left(\frac{\pi(x-a)}{b-a}\right)+\frac{1}{2} & x \in \left[a,b\right],\\ 
			0 & x \in \left]b, \pi\right]. 
		\end{cases}}
	\end{align}
\end{mdef}
\mpar{\label{rev:22}}\noindent\rev{Note that for~$ {a\neq b} $ this function is~$ \mathcal{C}^1 $ and smoothly transitions from~$ 1 $ to~$ 0 $.
In particular, both boundary values are included, i.e., points can be given both weights~$ 1 $ and~$ 0 $, which corresponds to fully taking them into account or to not taking them into account at all.
The threshold parameter~$ {a\in[0,\rev{\pi}]} $ translates the curve along the \mbox{$x$-axis} and controls where the cosine curve starts.
Similarly, the threshold parameter~$ {b\in[a,\pi]} $ controls where the cosine curve ends, i.e., the curve's decline is controlled by the distance between these two thresholds.
In particular, when choosing~$ {a=b=\pi} $, all inputs obtain uniform weight~$ 1 $ while for~$ {a=b} $, the function models a sharp cut-off at the chosen threshold.
This allows us to relate our weights to the uniform weights used in~\cite{weinmann2014semantic} and to the sharp cut-off weights of~\cite{yadav2018constraint}, respectively.}

\section{Evaluation Model}
\label{sec:Evaluation}

\noindent Having presented the set of neighborhood weights in Equation~(\ref{equ:Weights}) and the corresponding weighting function in Equation~(\ref{equ:sigmoid_cos}) in the previous section, we will now describe the mathematical background of our evaluation process.
For this, we turn to the information measures originally introduced by Shannon~\cite{shannon1948mathematical}.
Specifically, we will use a variation of the quantities derived in~\cite{demantke2011dimensionality,weinmann2014semantic} as we will present in Section~\ref{sec:NonDegenerateCovarianceMatrix}. First, we will establish the necessary notation and preliminary results.

Consider the covariance matrices~$ {C_i \in \mathbb{R}^{3 \times 3}} $ given by
\begin{align}
	\label{equ:CovarianceMatrix}
	C_i := \sum_{j\in\mathcal{N}_i} w_{ij} (p_j-\bar{p}_i)(p_j-\bar{p}_i)^T,
\end{align}
with ${i \in [n]}$, where~$\bar{p}_i=\frac{1}{|\mathcal{N}_i|}\sum_{j\in\mathcal{N}_i}p_j$ is the barycenter of the neighborhood of~$p_i$, \mpar{\label{rev:25}}\rev{thus $ (p_j-\bar{p}_i) $ is a column vector in $\mathbb{R}^3$ and~$(p_j-\bar{p}_i)^T\in\mathbb{R}^{1\times 3}$ its transpose, i.e., a row-vector.}
The weights~$ w_{ij} $ are chosen according to Equation~(\ref{equ:Weights}). The covariance matrix~$C_i$ is symmetric and positive-semi-definite. Thus, it has three non-negative eigenvalues, which in the following we will denote by
\begin{align}
	\label{equ:nonZeroEV}
	\lambda_i^1\geq\lambda_i^2\geq\lambda_i^3 \geq 0.
\end{align}
Depending on the neighborhood~$\mathcal{N}_i$ and the assigned weights~$w_{ij}$, we can prove the following theorem about the covariance matrix~$C_i$.
\begin{prop}[Non-degenerate Covariance Matrix]
	\label{prop:NonDegenerateCovarianceMatrix}
	For a set of points~$ {P=\{p_i\mid i\in[n]\}} $, fix a point~$ {p_i \in P} $ and its neighborhood~$ {\mathcal{N}_i\subseteq[n]} $, and consider the function~$ \sig^{\cos}_{a, b} $ from Equation~(\ref{equ:sigmoid_cos}) as well as the covariance matrix~$ C_i $ given in Equation~(\ref{equ:CovarianceMatrix}). 
	Assume there are~$ {\ell_1, \ell_2 \in \mathcal{N}_i} $, $ {\ell_1 \neq \ell_2} $ such that~$ {p_{\ell_1} \neq p_{\ell_2}} $ and~$ {n_{\ell_1} \neq -n_{\ell_2}} $. 
	Then, for some $ {b \in [a, \pi]} $, the sum of all eigenvalues of~$ C_i $ is strictly positive, independent of the choice of~$ {a \in [0,\pi]} $.
\end{prop}
\noindent \rev{\mpar{\label{rev:15}}Note, that a non-degenerate covariance matrix can trivially be obtained by setting $ a = \pi $. However, the proposition makes an even stronger statement, namely that degeneracy can be obtained independent from the choice of $a$. 
Its proof follows from the observation that the weights~$ w_{ij} $ are non-negative, as are all eigenvalues of~$ C_i $ since $C_i$ is positive semi-definite.
Thus, the sum of the eigenvalues is~$ 0 $ if and only if all eigenvalues are.
By a case distinction on the zero set of the function~$ \sig^{\cos}_{a, b} $, we can then prove that there exists some~$ {b \in [a, \pi]} $ which results in strictly positive weights, which proves the proposition.}

\subsection{Non-Degenerate Covariance Matrix}
\label{sec:NonDegenerateCovarianceMatrix}

\begin{figure}
	\centering
	\begin{tikzpicture}
	\pgfplotsset{		
		samples=150,
		axis lines=left,
	}
	\begin{axis}[
	domain=0:1, xmax=1.1,
	restrict y to domain=0:0.5, ymax=0.41,
	y=10cm,
	x=5cm,
	xlabel={$x$},
	ylabel={$-x\ln(x)$}
	]
	\addplot [db, thick] {-x*ln(x)};
	\end{axis}
	\end{tikzpicture}
	\caption{Plot of the summand~$-x\ln(x)$ from Equation~(\ref{equ:Edim}) for~$x\in[0,1]$ as all arguments~$L_i^\lambda$, $P_i^\lambda$, $S_i^\lambda$, and $ {\frac{\lambda_i^l}{\lambda_i^{\Sigma}} \forall l \in [1, 2, 3]} $ are taken from~$[0,1]$.}
	\label{fig:xln(x)Plot}
\end{figure}
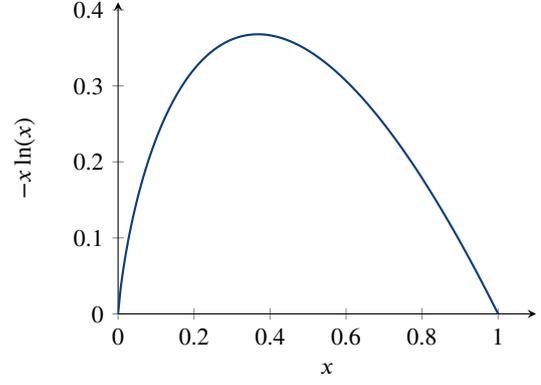

\noindent Given the assumptions of Proposition~\ref{prop:NonDegenerateCovarianceMatrix}, we can assume that $ {C_i\neq0\in\mathbb{R}^{3\times3}} $. Therefore, we can derive certain quantities from the eigenvalues of the covariance matrix.
In our context, we will consider the linearity~$L_\lambda$, planarity~$P_\lambda$, and scattering~$S_\lambda$.
These are given by
\begin{align}
	\label{equ:LinPlanScat}
	L_i^\lambda=\frac{\lambda_i^1-\lambda_i^2}{\lambda_i^1}, && P_i^\lambda=\frac{\lambda_i^2-\lambda_i^3}{\lambda_i^1}, && S_i^\lambda=\frac{\lambda_i^3}{\lambda_i^1}
\end{align}
and represent 1D, 2D, and 3D features in the point set, respectively.
See~\cite{demantke2011dimensionality} for a derivation and a detailed explanation of these quantities.
As~$ {C_i\neq0} $, we have~$ {\lambda_i^1\neq0} $, therefore the quantities in Equation~(\ref{equ:LinPlanScat}) are well-defined. 
Furthermore, because of the ordering of the eigenvalues given in Equation~(\ref{equ:nonZeroEV}), we have~$ {L_i^\lambda,P_i^\lambda,S_i^\lambda\in[0,1]} $. 
Hence, as 
\begin{align*}
	L_i^\lambda+P_i^\lambda+S_i^\lambda=1,
\end{align*}
each of these three quantities can be interpreted as the probability of the considered point to be part of an intrinsic 1D, 2D, or 3D part of the geometry.
The authors of~\cite{demantke2011dimensionality,weinmann2014semantic} consider the error
\begin{align}
	\label{equ:Edim}
	E_i^{\dim} = -L_i^\lambda\ln(L_i^\lambda)-P_i^\lambda\ln(P_i^\lambda)-S_i^\lambda\ln(S_i^\lambda).
\end{align}
See Figure~\ref{fig:xln(x)Plot} for a plot of each summand of the equation. 
Note that while~$ {\lim_{x\rightarrow0}\ln(x)=\infty} $ it is~$ {\lim_{x\rightarrow0}x\ln(x)=0} $, \rev{\mpar{\label{rev:16}}which follows from rewriting it as quotient and applying L'Hôpital's rule}. 
Practically, the error measure~$ E_i^{\dim} $ assesses to what extent the neighborhood~$\mathcal{N}_i$ indicates a corner, an edge point, or a planar point of the geometry. 
In particular, the extreme cases
\begin{align}
	\label{equ:ExtremalEigenvalues}
	(\lambda_i^1,\lambda_i^2,\lambda_i^3) \in \{ (\rho,0,0), (\rho,\rho,0), (\rho,\rho,\rho) \mid \rho\in\mathbb{R}_{>0}\}
\end{align}
all obtain~$E_i^{\dim}=0$.
\rev{
	That is to say that if a point~$p_i$ can be clearly classified as part of a linear, planar, or scattered segment of the point cloud, the classification error $E_i^{\dim}$ will indicate this.
}

\mpar{\label{rev:11}} \rev{
Note that in general applications, these extreme cases are unlikely to occur.
In particular in the presence of noise, the quantities~$ L_i^\lambda $, $ P_i^\lambda $, and~$ S_i^\lambda $ will generally not satisfy Equation~(\ref{equ:ExtremalEigenvalues}).
Thus, the classification error~$ E_i^{\dim} $ will be larger and therefore indicate that the point could not clearly be classified as part of a linear, planar or scattered segment of the point cloud.
}

\rev{We will use the classification error~(\ref{equ:Edim}) in our quantitative experiments in Section~\ref{sec:EvaluationResults}.
Aiming for as-clear-as-possible classification of points, we pursue as-small-as-possible values of~$ E^{\dim} $.	
However, the above discussion depends on the assumptions provided in Proposition~\ref{prop:NonDegenerateCovarianceMatrix}.} 
In the following we will discuss cases in which these assumptions are not satisfied.

\subsection{Degenerate Covariance Matrix}
\label{sec:DegenerateCovarianceMatrix}

\noindent In practical applications, the assumptions of Proposition~\ref{prop:NonDegenerateCovarianceMatrix} are not always satisfied. Note here that the classification error~$E_i^{\dim}$ is evaluated on a single point~$p_i$ of the point set~$P$. The following reasons can hinder the correct evaluation:
\begin{itemize}
	\item[\textit{i)}] If the point set contains multiple duplicates of a point, more than the sought-for number of neighbors~$k$, all points in the reported neighborhood collapse into a single point equal to the barycenter of the neighborhood. Thus,~$C_{i}$ becomes~$0$. 
	\item[\textit{ii)}] If a point~$p_i$ has a flipped normal in comparison to all its neighboring points~$p_j$, the argument~$x$ in the weight equation~$ {w_{ij}=\sig_{a,b}^{\cos}(x)} $ becomes~$\pi$ and therefore, all weights degenerate to~$0$. This happens in particular for very small or thin geometries as well as for faulty normal fields.
	\item[\textit{iii)}] Even if the assumptions of Proposition~\ref{prop:NonDegenerateCovarianceMatrix} are satisfied, it only states the existence of a suitable parameter~$ {b\in[a,\pi]} $. Therefore, choosing parameter~$b$ too small can cause all weights in the covariance matrix~(\ref{equ:CovarianceMatrix}) to degenerate to~$0$.
\end{itemize}

\noindent In the following evaluation, we prevent case~\textit{i)} by requiring the point sets to only contain distinct points.
Furthermore, we orient the normal field to prevent case~\textit{ii)}. Concerning a too small parameter~$b$, we report a failure in the computation of the error values for the point~$p_i$ if~$ {\sum_{\ell=1}^{3}\lambda_i^\ell=0} $. 
By including the choice~$a=\pi$ for the parameters, we ensure that each model has at least one correctly evaluated error value~$ E_i^{\dim} $ at each point~$ {p_i\in P} $.


\section{Evaluation Results}
\label{sec:EvaluationResults}

\noindent In this section, we present our quantitative evaluation of the weights presented in Equation~(\ref{equ:Weights}).
For the evaluation, we utilize the classification error~$E^{\dim}$ as defined in Equation~(\ref{equ:Edim}).
Our clean models are taken from a data set described in~\cite{hu2018tetrahedral}.
The authors provide ten thousand clean and manifold surface meshes, which are obtained by exporting only the boundary of the tetrahedral meshes used in~\cite{hu2018tetrahedral}.
From these, we randomly select a subset of~$1,000$ meshes with uniform probability.
Furthermore, we use~$100$ meshed models each from the real-world object scans provided by~\cite{choi2016large} \rev{and by~\cite{bogo2014faust}.
Finally, to test the scalability of our approach, we also include the model ``Pan et Oursons'' from~\cite{laric2012three}.}

\begin{figure*}
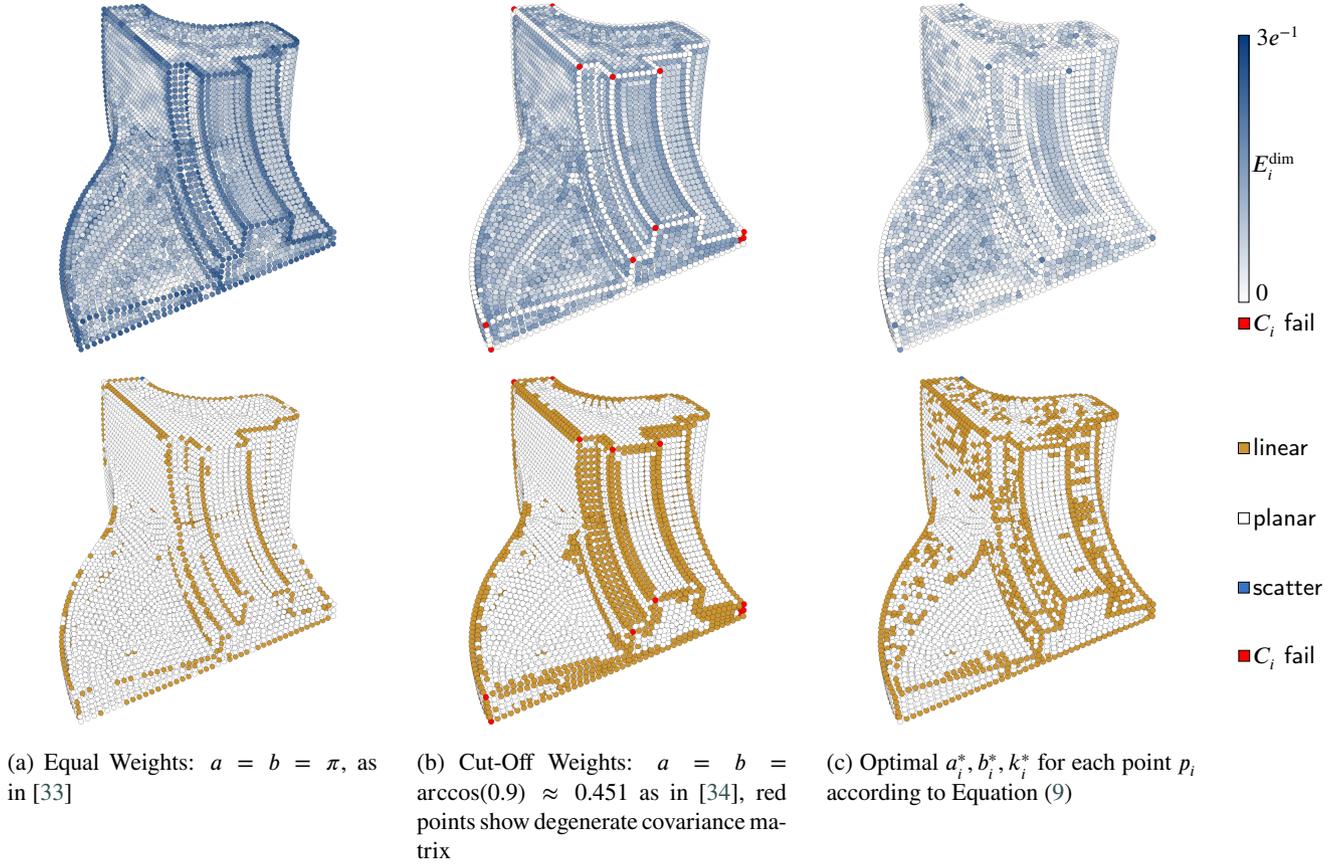

	\centering
	\begin{subfigure}[t]{0.28\textwidth}
		\includegraphics[width=1.\textwidth]{img/error-LPS-Plots/fandisk_equalWts_error}\\
		\includegraphics[width=1.\textwidth]{img/error-LPS-Plots/fandisk_equalWts_lps}
		\caption{Equal Weights: $a=b=\pi$, as in~\cite{weinmann2014semantic}}
		\label{fig:fandiskEqual}
	\end{subfigure}
	\hfill
	\begin{subfigure}[t]{0.28\textwidth}
		\includegraphics[width=1.\textwidth]{img/error-LPS-Plots/fandisk_cutoffWts_error}\\
		\includegraphics[width=1.\textwidth]{img/error-LPS-Plots/fandisk_cutoffWts_lps}
		\caption{Cut-Off Weights: $a=b=\arccos(0.9)\approx0.451$ as in~\cite{yadav2018constraint}, red points show degenerate covariance matrix}
		\label{fig:fandiskCutOff}
	\end{subfigure}
	\hfill
	\begin{subfigure}[t]{0.28\textwidth}
		\includegraphics[width=1.\textwidth]{img/error-LPS-Plots/fandisk_allWts_error}\\
		\includegraphics[width=1.\textwidth]{img/error-LPS-Plots/fandisk_allWts_lps}
		\caption{Optimal $a_i^*,b_i^*,k_i^*$ for each point~$p_i$ according to Equation~(\ref{equ:OptimalParameterPairs})}
		\label{fig:fandiskOptimal}
	\end{subfigure}
	\hfill
	\begin{subfigure}[t]{0.07\textwidth}
		\def\svgwidth{1.\textwidth}
		\\[0.5cm]
		\def\svgwidth{1.\textwidth}
		
	\end{subfigure}
	\caption{
		The effect of the different parameters on the fandisk model. 
		The top row shows the classification error~$E_i^{\dim}$ from Equation~(\ref{equ:Edim}) for each point of the model, from low (white) to high error (dark blue). 
		Note how the optimal weights from Equation~(\ref{equ:OptimalParameterPairs}) have drastically reduced error in comparison to both equal weights (used by~\cite{weinmann2014semantic}) and sharp cut-off weights (used by~\cite{yadav2018constraint}).
		The red points indicate elements of the point set, for which the covariance matrix from Equation~(\ref{equ:CovarianceMatrix}) degenerates given the chosen weights.\\
		Bottom row shows a feature classification according maximum value out of linearity (orange), planarity (white), and scattering (dark blue) as defined in Equation~(\ref{equ:LinPlanScat}).
		Note how equal weights fail to consistently identify edge structures.
		The cut-off weights manage to identify planar areas well while over-pronouncing edge structures.
		These are identified well by the weights from Equation~(\ref{equ:OptimalParameterPairs}).
	}
	\label{fig:fandiskErrors}
\end{figure*}

\rev{For all these models}, we use the mesh information and its manifold property to obtain oriented face normals. 
From these, we compute vertex normals and then use these and the vertices as point sets for our experiments.
For each such point set~$P$, we consider the parameter sets
\begin{align*}
	\rev{\mathfrak{A} := \mathfrak{B} := \left\{0,\frac{\pi}{6},\frac{\pi}{3},\frac{\pi}{2},\frac{2\pi}{3},\frac{5\pi}{6},\pi\right\}}
\end{align*}
\mpar{\label{rev:27}}\rev{for the choice of $a$ and $b$, respectively, where we ensure that $ {a\leq b} $.
We choose this range as a reasonable trade-off between complexity of the experiments and exploration of the parameter space.
Note that in Section~\ref{sec:ApplicationScenario}, we consider further parameter values that are rooted in the application scenario considered there.}
We use the combinatorial neighborhood notion\footnote{For a point~$ p_i\in P$, we consider the index~$ i $ as well as the indices of the~$ k $ nearest neighbors to~$ p_i $ within~$ P $ as neighborhood~$ \mathcal{N}_i $, i.e.,~$ {|\mathcal{N}_i|=k+1} $.}, so that for every pair~$ \rev{(a, b)\in\mathfrak{A}\times\mathfrak{B}} $ and every point~$ {p_i\in P} $, we calculate its $ E_i^{\dim} $ value over the range of~$ k $, taken from
\begin{align*}
	\mathfrak{K} := \lbrace 6, \ldots, 20 \rbrace.
\end{align*}
\mpar{\label{rev:27b}}\rev{We assume this range for~$ k $, as it reflects typical, heuristic choices for neighborhood sizes in the area of point set processing, see the works discussed in Section~\ref{sec:RelatedWork}, in particular~\cite{pauly2003shape}.} 
\rev{\mpar{\label{rev:18}}For each point~$ p_i $ in each point set~$P$, we obtain an optimal parameter triple~$ {\left(a_i^*,b_i^*,k_i^*\right)} $ as
\begin{align}
	\label{equ:OptimalParameterPairs}
	\left(a_i^*,b_i^*,k_i^*\right) &= \argmin_{\left(a,b,k\right)\in\mathfrak{A}\times\mathfrak{B}\times\mathfrak{K}} E_{i}^{\dim}.
\end{align}
Following the discussion from Section~\ref{sec:DegenerateCovarianceMatrix}, we set~$ {E_i^{\dim}=\infty} $ if the covariance matrix~$ C_i $ for the point~$ {p_i\in P} $ degenerates for all parameter choices~$ {(a,b,k)\in\mathfrak{A}\times\mathfrak{B}\times\mathfrak{K}} $.}

\mpar{\label{rev:13}}\rev{See Figure~\ref{fig:fandiskErrors} for an illustration of the classification error~$E^{\dim}$ on the fandisk geometry as well as for a comparison of different parameter choices~$(a,b)$. 
The top row shows the classification error~$ E_i^{\dim} $ from Equation~(\ref{equ:Edim}) on each point of the geometry, colored from low error (white) to large error (dark blue).
Note that when fixing parameters~$ {(a,b)} $, it is possible that the covariance matrix~$ C_i $ degenerates for every choice~$ {k\in\mathfrak{K}} $.
This happens for the specific choice $ {a=b=\arccos(0.9)} $ as used in~\cite{yadav2018constraint}.
We have colored the respective points red.
The optimal triple from Equation~(\ref{equ:OptimalParameterPairs}) achieves significantly lower classification error~$E_i^{\dim}$ than the equal weights of~\cite{weinmann2014semantic} or the cut-off weights of~\cite{yadav2018constraint}.
The observed fluctuation in planar areas is due to (a) the utilization of combinatorial neighborhoods, which do not always provide symmetrically shaped neighborhoods on a synthetic geometry like the fandisk, as well as to (b) the sensitivity of~$E_i^{\dim}$ to slight changes in the covariance matrix.}

\rev{The bottom row of Figure~\ref{fig:fandiskErrors} shows a feature classification according to the maximum value out of linearity (orange), planarity (white), and scattering (dark blue) as defined in Equation~(\ref{equ:LinPlanScat}).
Note how the equal weights of~\cite{weinmann2014semantic} classify almost all elements as planar and fail to identify edge structures.
In contrast, the cut-off weights of~\cite{yadav2018constraint} identify all edges, but over-pronounce them.
The optimal weight choice from~(\ref{equ:OptimalParameterPairs}) takes a middle ground between these two extremes, on the cost of identifying several clearly planar points as linear.
Again, this stems to a certain extend from the run a synthetic geometry.
Observe that the equal weights and the optimal weights do identify scattered points (one example being the topmost corner of fandisk) while the cut-off weights rather fail to create a covariance matrix at corner points.
}

\rev{
	The images in Figure~\ref{fig:fandiskErrors} summarize our following experiments.
	In order to compare with the findings of~\cite{weinmann2014semantic}, we compute the classification error~$ E_i^{\dim} $ for each point of every point set of the three chosen model repositories~\cite{hu2018tetrahedral,bogo2014faust,choi2016large} as well as of the single, large model from~\cite{laric2012three}.	
}
In the following we report and interpret our findings.


\begin{figure*}
	\begin{subfigure}[t]{0.49\textwidth}
		\def\svgwidth{1.\textwidth}
		{\tiny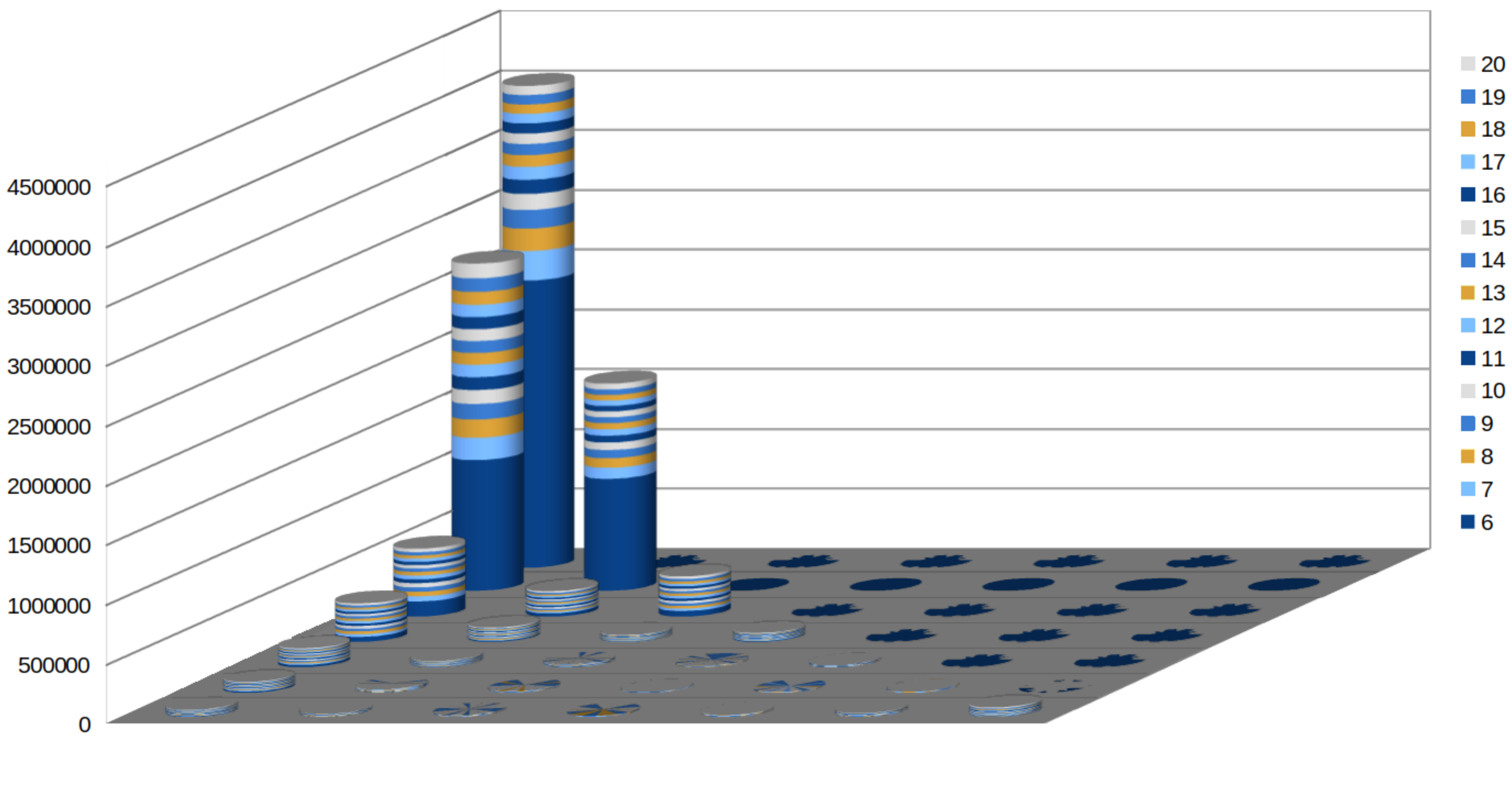}
		\caption{Applied to~$ 1,000 $ geometries randomly selected from the data set used in~\cite{hu2018tetrahedral}, with $7,213,429$ total points.}
		\label{fig:panABHist}
	\end{subfigure}
	\hfill
	\begin{subfigure}[t]{0.49\textwidth}
		\def\svgwidth{1.\textwidth}
		{\tiny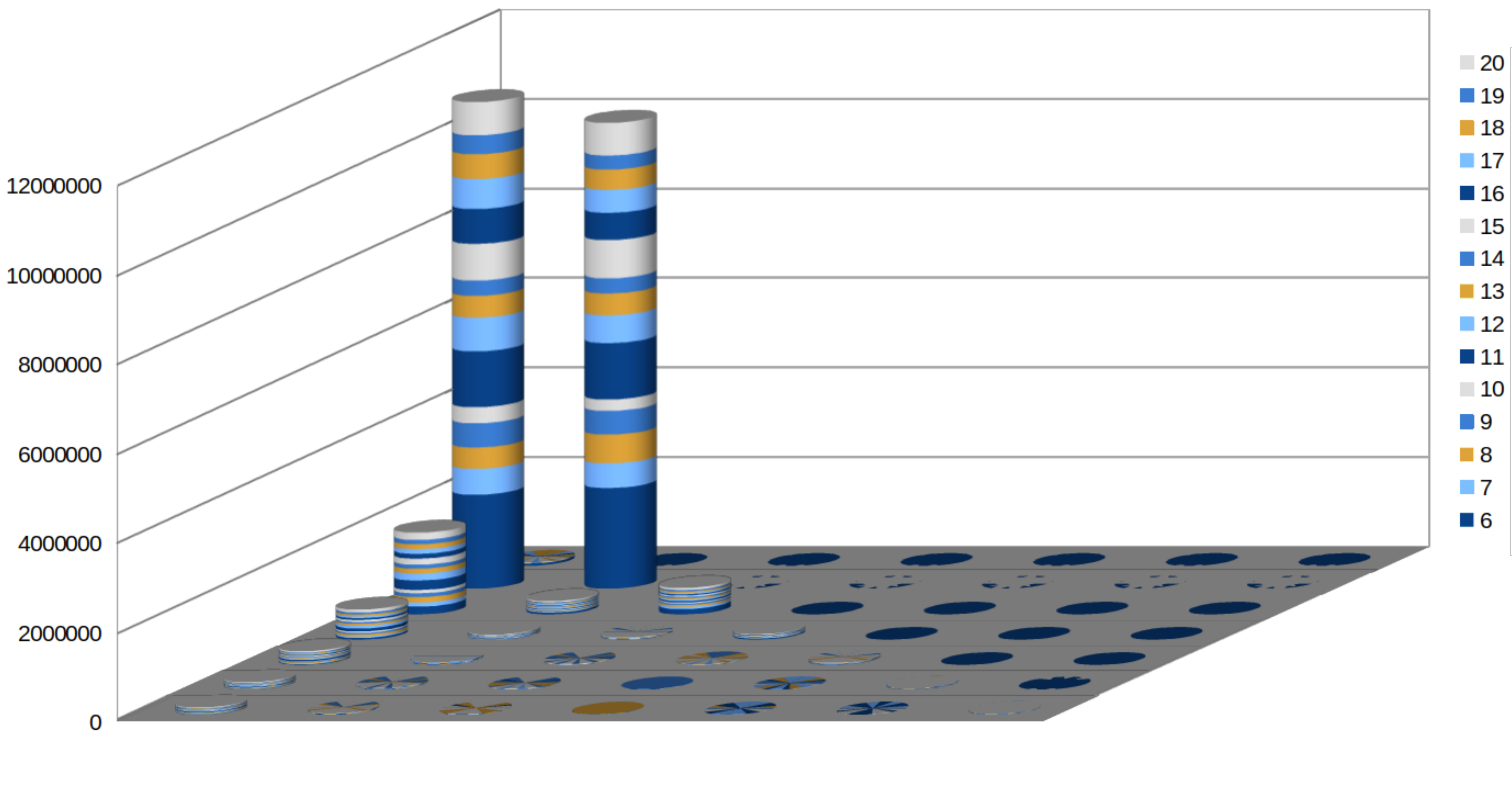}
		\caption{Applied to~$ 100 $ geometries taken from~\cite{choi2016large}, with $25,929,256$ total points.}
		\label{fig:choiABHist}
	\end{subfigure}\\
	\begin{subfigure}[t]{0.49\textwidth}
		\def\svgwidth{1.\textwidth}
		{\tiny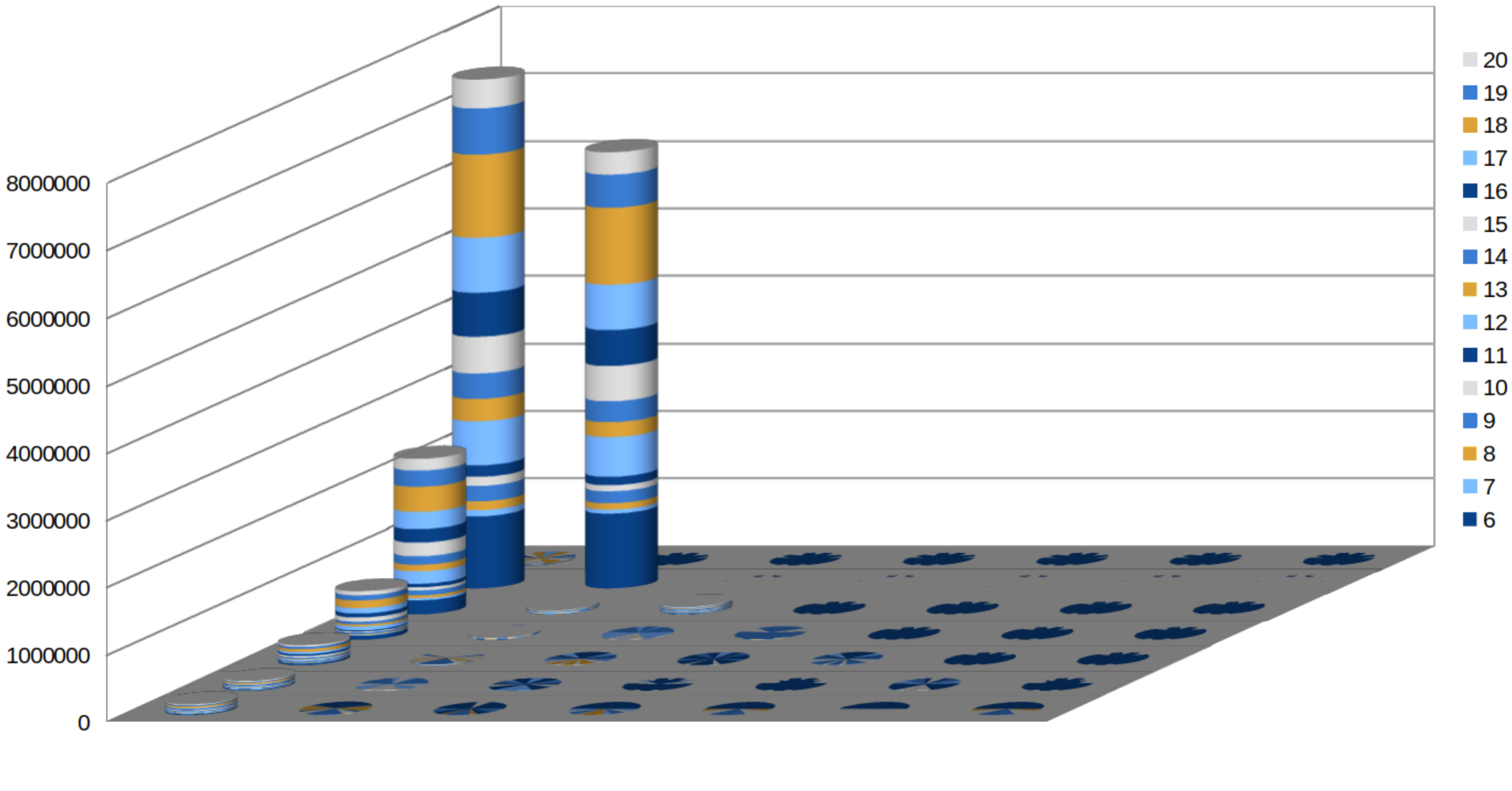}
		\caption{Applied to the~$ 100 $ geometries from the data set presented in~\cite{bogo2014faust}, with $17,918,016$ total points.}
		\label{fig:faustABHist}
	\end{subfigure}
	\hfill
	\begin{subfigure}[t]{0.49\textwidth}
		\def\svgwidth{1.\textwidth}
		{\tiny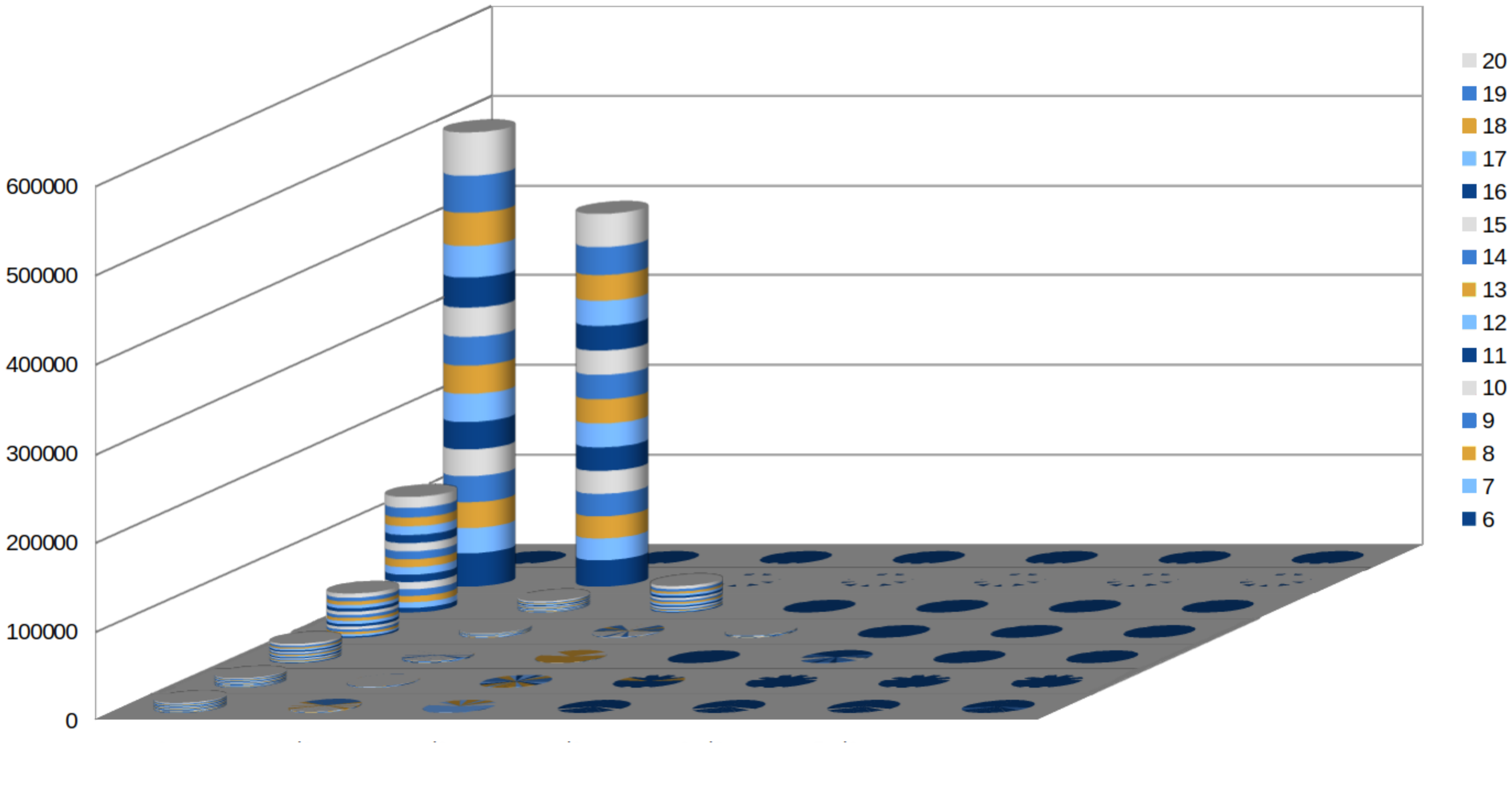}
		\caption{Applied to the ``Pan et Oursons'' scan from~\cite{laric2012three}, with $1,199,992$ total points.}
		\label{fig:threedscanABHist}
	\end{subfigure}
	\caption{
		Histograms of preferred sigmoid parameters~$ (a^*, b^*, k^*) $ (Eq.~\ref{equ:OptimalParameterPairs}) with respect to minimal error values for~$ E^{\dim} $ (Eq.~(\ref{equ:Edim})) over the range~$ \mathfrak{K} $ when applied to several large model repositories.
		Each point of the respective point set(s) corresponds to one unit in the histogram.
		Additionally, each such bar is colored according to the chosen optimal neighborhood size~$k$, from the lowest at the bottom to the highest at the top.
	}
	\label{fig:Histograms}
\end{figure*}

\subsection{Global \rev{$(a, b, k)$} Analysis}
\label{sec:GlobalABAnalysis}

\noindent We analyze the total amount of $ \rev{(a, b, k)} $ choices for all model repository selections. 
Here, we count all points of all point sets with their respective optimal parameter triple~$ (a^*, b^*, k^*) $. 
The corresponding four global histograms for the three model repositories and the model from~\cite{laric2012three} are given in Figure~\ref{fig:Histograms}.
\rev{There, each point of each geometry contributes one unit in the histograms, which report the number of points that choose a parameter combination~$ (a,b) $, with $a$ on the $x$-axis and $b$ on the $y$-axis.
Additionally, each such bar is colored according to the chosen optimal neighborhood size~$k$, from the lowest at the bottom to the highest at the top.}
In summary, the classification error acts similar on all data sets, i.e., in the comparison between clean and real-world models.

On the large scale of~$ 1,000 $ point sets with a total of $ 7,213,429 $ points (Figure~\ref{fig:panABHist}), we observe, that on average, a small choice for parameter~$ a $ and a similarly choice for parameter~$ b $ are preferred. 
This can be interpreted to say that it is desirable to take only normals into account that exhibit a small deviation.
\rev{In particular, Figure~\ref{fig:panABHist} suggests that the majority of points from the clean models choose a neighborhood without any room for normal deviation $ {(a=b=0)} $. 
This is one notable difference to the histograms on scanned models, Figures~\ref{fig:choiABHist} to~\ref{fig:threedscanABHist}, where these drastic weights are almost never chosen.}

\rev{It is particularly noteworthy that almost no points chose equal weights $ {a=b=\pi} $ which highlights the benefit of our approach over that chosen by~\cite{weinmann2014semantic}.
Furthermore, choosing a sharp cut-off along the lines of~\cite{yadav2018constraint}, by $ a=b=\pi/6 $, occurs for about a quarter of the points.
However, about $38\%$ rather go with a softer decrease by choosing $a=0, b=\pi/6$.}
A localized, i.e., model-depended, discussion about the possibility to increase~$ a $ and~$ b $ for better results is given in the upcoming section.

\rev{In terms of scanned real-world models (Figures~\ref{fig:choiABHist} to~\ref{fig:threedscanABHist}), we analyzed~$100$ point sets from each~\cite{choi2016large,bogo2014faust} and one large model from~\cite{laric2012three}.
In comparison to the clean models, we do observe a different behavior. 
Namely, while small values for~$ a $ and $ b $ are still favored, the choice of $ a=0 $, which was most prominent on clean models, is almost never made for scanned models. 
The chosen weights indicate that mostly neighborhoods with a normal deviation of up to~$\pi/6$ are taken into account. These are either all weighted uniformly $(a=\pi/6)$ or with gradually deteriorating weights $(a=0)$.
We interpret the parameter~$ b $ to reflect the noise components caused by the acquisition process. 
Therefore, choosing the lowest possible choice of $b$ causes several points~$ {p_i\in P} $ to have degenerate covariance matrices~$C_i$, independent of the chosen neighborhood size~$k$.}
See also the following section for a more detailed discussion of this.

\rev{In conclusion, we see that weight-determination generally favors a narrow window between parameters $a$ and $b$.
This corresponds to using a neighborhood with an overall small normal deviation.
The value~$ b $ however depends on the geometry. 
Clean models mostly attain smaller error values for very small values of~$ b $, whereas real-world models require slightly larger of~$ b $ to obtain non-degenerate covariance matrices.
All models from all repositories have in common that they almost never report equal weights as preferred weight assignment.
Hence, when regarding the classification error~$ E_i^{\dim} $, the equal weighting scheme of~\cite{weinmann2014semantic} is inferior to the family of weights presented here.
This becomes obvious when comparing the values obtained from our experiments, see Table~\ref{tab:WeinmannComparison}.
The classification error computed with our weights~(\ref{equ:OptimalParameterPairs}) has lower minimum, average, maximum, and standard deviation than the error computed with the equal weights of~\cite{weinmann2014semantic}.
}

\begin{table*}
	\def\arraystretch{1.2}
	\begin{tabular}{r|l||l|l|l|l}
		&& $\min_i(E_i^{\dim})$ & $\frac{1}{N}\sum_{i=1}^N E_i^{\dim}$ & $\max_i(E_i^{\dim})$ & $\sd_i(E_i^{\dim})$ \\
		\hline
		\hline
		\textbf{Clean~\cite{hu2018tetrahedral}} & Weights~\cite{weinmann2014semantic} & 0 & 0.6309891 & 1.071584 & 0.1896867\\
		\hline
		$ 7,213,429 $ points & Our~(\ref{equ:OptimalParameterPairs}) & 0 & 0.2477968 & 1.012825 & 0.1882686\\
		\hline
		\hline
		\textbf{Scanned~\cite{choi2016large}} & Weights~\cite{weinmann2014semantic} & 2.946194$\cdot10^{-8}$ & 0.3348262 & 1.071796 & 0.1967012\\
		\hline
		25,929,256 points & Our~(\ref{equ:OptimalParameterPairs}) & 0 & 0.2526083 & 0.9649821 & 0.1569379\\
		\hline
		\hline
		\textbf{Scanned~\cite{bogo2014faust}} & Weights~\cite{weinmann2014semantic} & 0.001344446 & 0.2497551 & 1.048021 & 0.1144664 \\
		\hline
		17,918,016 points & Our~(\ref{equ:OptimalParameterPairs}) & 0 & 0.2109502 & 0.9379431 & 0.09631257 \\
		\hline
		\hline
		\textbf{Scanned~\cite{laric2012three}} & Weights~\cite{weinmann2014semantic} & 0.003978099 & 0.3827366 & 0.9805029 & 0.1456077 \\
		\hline
		1,199,992 points & Our~(\ref{equ:OptimalParameterPairs}) & 0 & 0.3200202 & 0.7774453 & 0.1337365 \\
	\end{tabular}
	\caption{Quantitative comparison of the classification error~$E_i^{\dim}$ computed over different model repositories with weights by~\cite{weinmann2014semantic} and our weights~(\ref{equ:OptimalParameterPairs}). 
		Note that our weighting scheme always obtains lower minimal, average, and maximum error as well as a lower standard deviation.}
	\label{tab:WeinmannComparison}
\end{table*}

Sharp cut-off weights are only chosen as optimal weighting by a subset of the real-world scans.
As~\cite{yadav2018constraint} used sharp cut-off weights in the context of denoising, our results hint that this weight set might be beneficial in the presence of noise.
However, for about~$75\%$ of the scanned models, when considering the classification error~$E_i^{\dim}$ our weighting family still chooses weights superior to the cut-off weights used by~\cite{yadav2018constraint}.


\subsection{Local (a,b) Analysis}
\label{sec:LocalABAnalysis}

\noindent \rev{In this section, we will discuss the~$ {(a^*, b^*)} $ choices presented in the previous section from a local, i.e., point-set-dependent, perspective. 
The respective results are presented in Table~\ref{tab:ABlocal}. 
There, the first row corresponds to the clean models from~\cite{hu2018tetrahedral} while the other three rows correspond to the scanned real-world models from~\cite{choi2016large,bogo2014faust,laric2012three}. 
The columns present information about the amount of points accepting minimal value $ {a=0} $, allowing $(a-)$ or forbidding $(\neg a-)$ a decrease of~$a$, accepting minimal value~$b=a$, and allowing $(b-)$ or forbidding $(\neg b-)$ a decrease of~$b$.
In this scheme, the columns $ \neg a- $ and $ \neg b- $ denote the percentage of those points for which a decrease of the respective parameter results in a degenerate covariance matrix~$ C_i $, see Section~\ref{sec:DegenerateCovarianceMatrix}. 
Observe that we cover all possible cases.
For easy comparability, we provide the respective case numbers in percent, with the total number of points for the respective repository given in the last column.}

\begin{table*}
	\def\arraystretch{1.2}
	\begin{center}
			\begin{tabular}{l||rrr|rrr|r}
				& $ a = 0 $ & $ a- $ & $ \neg a- $ & $ b = a $ & $ b- $ & $ \neg b- $ & \# Points\\
				\hline\hline
				\bf Clean \cite{hu2018tetrahedral}  & 60.19\% & 39.81\% & 0\% & 36.80\% & 25.49\% & 37.72\% & 7,213,429 \\
				\hline
				\bf Scanned	\cite{choi2016large} & 54.30\% & 45.70\% & 0\% & 43.50\% & 14.21\% & 42.29\% & 25,929,256\\
				\hline
				\bf Scanned \cite{bogo2014faust} & 62.44\% & 37.56\% & 0\% & 36.86\% & 21.03\% & 42.11\% & 17,918,016 \\
				\hline
				\bf Scanned \cite{laric2012three} & 60.70\% & 39.30\% & 0\% & 37.33\% & 20.46\% & 42.21\% & 1,199,992
		\end{tabular}
	\end{center}
	\caption{Distribution of~$ (a^*, b^*) $ choices into the three cases of (a) an attained minimum~($ {a = 0, b = a} $), (b) a possible decrease of the parameter without failure~($ {a-, b-} $), and (c) impossibility of decreasing the parameter because it would cause a degenerate covariance matrix~($ \neg a-, \neg b- $).}
	\label{tab:ABlocal}	
\end{table*}

\rev{Having all values in one chart, we directly observe the behavior assessed for parameter~$ a $ in the previous section. 
There, we stated that especially in the case of clean models, an as-small-as-possible value for~$ a $ is favorable over larger values for~$ a $. 
Indeed, Table~\ref{tab:ABlocal} confirms this statement, as almost\footnote{There are $<0.01\%$ for each scanned repository that would allow for a decrease, which is not shown here due to rounding.} none of the points allows for an decrease of parameter~$a$ (cf.\@ column~$\neg a-$).
This justifies the small values for~$ a $ attained in the real-world scenarios presented in Figures~\ref{fig:choiABHist} to~\ref{fig:threedscanABHist} when compared to the values of~$ a $ attained in the clean scenarios in Figure~\ref{fig:panABHist}.
Semantically, this opts for including just enough neighbors in the computation to make it feasible, i.e., to prevent a degenerate covariance matrix, but focus on those that are as similar as possible with regard to the normal field.}

\rev{The reported numbers on the parameter~$ b $ also support the observation drawn before.
It is chosen to be as small as possible, i.e., as close to the chosen $a$ without creating a degenerate covariance matrix.
Over all repositories, $b$ is chosen to create a sharp cutoff ($b=a$) in about $36\%$ of the considered points.
A notable exception is the scanned data set~\cite{choi2016large}, which allows for $43.5\%$ of the points to choose a sharp cutoff.
This is possibly due to lower noise levels and different geometry types in this data set when compared to the other scanned data sets~\cite{bogo2014faust,laric2012three}.
Furthermore, in about $42\%$ of the scanned points and $37\%$ of the clean points, $b$ is at least chosen to be as-close-as-possible to $a$, i.e., the weighting scheme is chosen to be as-close-as-possible to a sharp cut-off, which cannot be realized because a further decrease of $b$ would cause a degenerate covariance matrix.
These observations justify the general weighting choice of sharp cutoff, as chosen by~\cite{yadav2018constraint}, although the particular chosen values only prove to be most effective in about one fourth of all models from the data set used here.}

Summarizing the global and local analysis of the parameter choices~$ (a^*,b^*) $, we draw the following conclusions:
\begin{itemize}
	\item \rev{The utilized classification error favors weight determination with as-small-as-possible values for both parameters~$ a $ and~$ b $. 
	That is, only points with as-similar-as-possible normals are considered, but out of these, all are allowed to influence the computation as evenly as possible.}
	\item \rev{Equal weights ($a=b=\pi$), as used by~\cite{weinmann2014semantic}, are never chosen as optimal parameters to obtain a minimal classification error~$E_i^{\dim}$.}
	\item \rev{Sharp cut-off weights as widely used in the literature, e.g., in~\cite{yadav2018constraint}, attain minimal classification error for $36.8\%$ of the clean points and for up to $43.5\%$ of the scanned points.
	This proves their relevance in particular for real-world scenarios.}
\end{itemize}


\subsection{Global k Analysis}

\begin{figure*}
	\centering
	\begin{tikzpicture}
	\begin{axis}[
	symbolic x coords = {6,7,8,9,10,11,12,13,14,15,16,17,18,19,20},
	xtick = {6,7,8,9,10,11,12,13,14,15,16,17,18,19,20},
	xlabel={Neighborhood Size $k$}, 
	ylabel={Points w/ neighborhood size $k$ $[\%]$},
	axis lines*=left, 
	ybar=0pt, 
	bar width=6pt, 
	height=0.35\textwidth, width=1.\textwidth, 
	legend style={at={(1.,1.)},anchor=north east},
	enlarge x limits=0.05,
	enlarge y limits = 0,
	]
	\addplot[fill=b, draw=b] coordinates {(6,0.35829243) (7,0.06387129) (8,0.05558854) (9,0.05056818) (10,0.04724022) (11,0.04482750) (12,0.04324518) (13,0.04211076) (14,0.04137880) (15,0.04097081) (16,0.04071143) (17,0.04062146) (18,0.04146696) (19,0.04303723) (20,0.04606921)};

	\addplot[fill=o, draw=db] coordinates {(6,0.18754055) (7,0.05357161) (8,0.05570368) (9,0.05039902)
	(10,0.03144564) (11,0.11281504) (12,0.06689070) (13,0.04861096) (14,0.03530047) (15,0.07670147) (16,0.06361363) (17,0.05547170) (18,0.04984728) (19,0.03921165) (20,0.07287660)};

	\addplot[fill=db, draw=db] coordinates {(6,0.14038920) (7,0.01204079) (8,0.01707946) (9,0.02956516)
		(10,0.01719945) (11,0.02193759) (12,0.08660713) (13,0.04000789) (14,0.05079552) (15,0.07853582) (16,0.08448982) (17,0.10660215) (18,0.16647686) (19,0.08849897) (20,0.05977420)};
	
	\addplot[fill=gray, draw=db] coordinates {(6,0.06380543) (7,0.05384203) (8,0.05597121) (9,0.05806622)
		(10,0.05997873) (11,0.06184125) (12,0.06343209) (13,0.06457293) (14,0.06571460) (15,0.06672961) (16,0.06799795) (17,0.07037297) (18,0.07329466) (19,0.08046137) (20,0.09391896)};

	\legend{
		clean models~\cite{hu2018tetrahedral}, 
		scanned models~\cite{choi2016large},
		scanned models~\cite{bogo2014faust},
		scanned models~\cite{laric2012three}
	}
	\end{axis}
	\end{tikzpicture}
	\caption{\rev{Histogram of preferred neighborhood sizes $ k^* $ with respect to minimal error value~$ E_i^{\dim} $.} To ensure compatibility over the different data repositories, we normalize by the total number of points and report the percentage of points choosing the respective neighborhood size.}
	\label{fig:kHist}
\end{figure*}
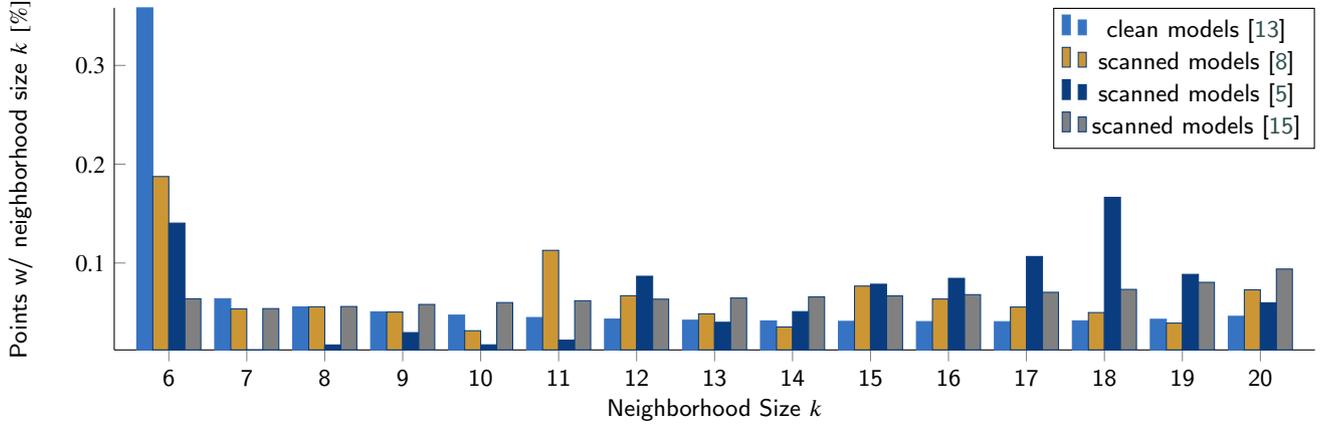

\noindent As stated in the beginning of Section~\ref{sec:EvaluationResults}, for each point in the utilized point sets, we \rev{also obtain a preferred neighborhood size~$ {k^*\in\mathfrak{K}} $ yielding smallest classification error $ E_i^{\dim} $ among all choices (Equation~(\ref{equ:OptimalParameterPairs})).} 
In Figure~\ref{fig:kHist}, we present a histogram plotting this data, i.e., for each neighborhood size $ {k\in\mathfrak{K}} $, we show what percentage of points \rev{from the respective model repository} use this~$ k $.

Note that the plot for \rev{the clean models shows a favor for} an as-small-as-possible neighborhood size~$ k $ over larger neighborhoods. 
\rev{In contrast, the scanned models show a different behavior. 
	Whereas the repository \cite{choi2016large} also attains its peak at $ k = 6 $, it is more equally distributed among the whole range, with a notable second peak at $ k = 11 $. 
	The models in \cite{bogo2014faust}, however, exhibit an almost Gaussian bump around their maximal value $ k = 18 $.
	Finally, the results for the model chosen from~\cite{laric2012three} are almost uniform over the entire range, with a slight increase for growing values of $ k $.}

\mpar{\label{rev:38}}\rev{In order to investigate scalability effects, we have included one significantly larger model from~\cite{laric2012three} in our analyses.
Note that the general observations as made on Figure~\ref{fig:Histograms}, Table~\ref{tab:WeinmannComparison}, Table~\ref{tab:ABlocal}, and Figure~\ref{fig:kHist} particularly hold for this model.
Furthermore, despite the fact that~\cite{weinmann2014semantic} focuses on these large-scale models, our weighting approach~(\ref{equ:OptimalParameterPairs}) still provides smaller classification errors as indicated in Table~\ref{tab:WeinmannComparison}.
Thus, we cannot report any scaling issues.}

\rev{For the clean models, we obtain an average neighborhood size of~$\bar{k}=10.55$ with a standard deviation of $ \sigma=4.77 $. For the scanned models, those quantities are:}
\begin{itemize}
	\item \rev{$ \bar{k}=12.09489$, $ \sigma=4.618431 $ (\cite{choi2016large}),}
	\item \rev{$ \bar{k}=14.22883$, $\sigma=4.487704 $ (\cite{bogo2014faust}),}
	\item \rev{and $ \bar{k}=13.54247$, $\sigma=4.398983 $ (\cite{laric2012three}).}
\end{itemize}
These findings suggest that variable neighborhood sizes yield smaller error values \rev{with regard to the classification error $E_i^{\dim}$}. 
\rev{If a global neighborhood size has to be chosen, then the average values provided by this analysis serve as reasonable choices, as they provide a good trade-off between a fixed neighborhood size and a low classification error.} 
In order to further investigate the benefit of varying neighborhood sizes, in the following section, we turn to a point-set-dependent perspective.


\subsection{Local $ k $ Analysis}
\label{sec:LocalKAnalysis}

\noindent We will now consider the standard variation of the neighborhood sizes taken over a single model for~$ E^{\dim} $. 
We aim to better understand and investigate the hypothesis formulated above, i.e., the statement that a variable neighborhood size contributes to lower classification error.

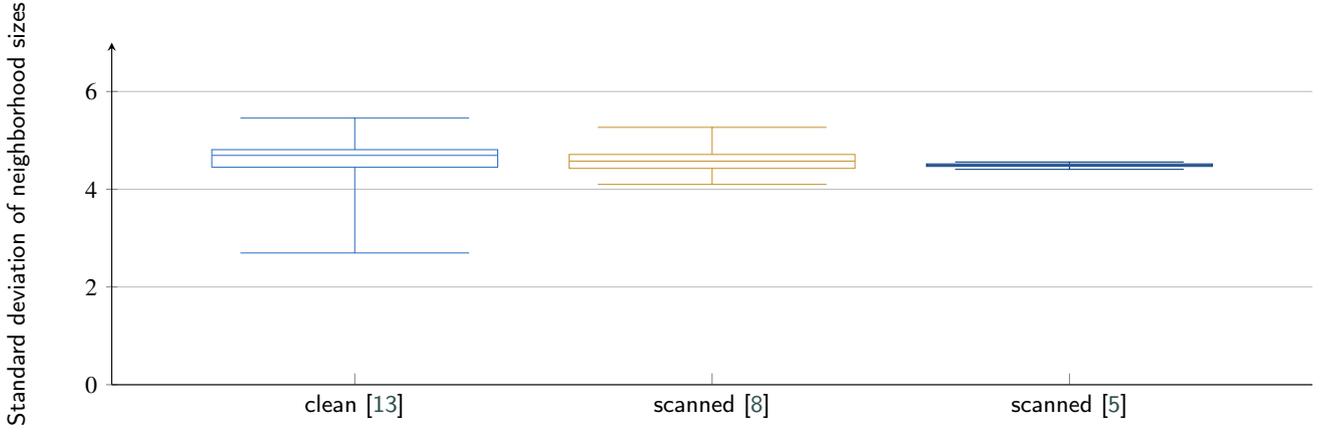
\begin{figure*}
	\centering
	\begin{tikzpicture}
	\begin{axis}[
	height=0.35\textwidth, width=1.\textwidth, 
	boxplot/draw direction=y,
	ylabel={Standard deviation of neighborhood sizes},
	axis x line*=bottom,
	axis y line=left,
	ymajorgrids,
	xtick={1,2,3,4,5,6},
	xticklabels={
		clean \cite{hu2018tetrahedral}, 
		scanned \cite{choi2016large}, 
		scanned \cite{bogo2014faust},
		scanned \cite{laric2012three}
	},
	cycle list={{b},{o},{db}},
	ymin = 0,
	ymax = 7,
	]
	\addplot+ [
	boxplot prepared={
		lower whisker=2.696194,
		lower quartile=4.450528729,
		median=4.693041,
		upper quartile=4.810183,
		upper whisker=5.456096,
	}
	] coordinates {};
	\addplot+ [
	boxplot prepared={
		lower whisker=4.099657,
		lower quartile=4.427528,
		median=4.574536,
		upper quartile=4.712061,
		upper whisker=5.266856,
	}
	] coordinates {};
	\addplot+ [
	boxplot prepared={
		lower whisker=4.407243,
		lower quartile=4.468007,
		median=4.489193,
		upper quartile=4.518713,
		upper whisker=4.557793,
	}
	] coordinates {};
	\end{axis}
	\end{tikzpicture}
	\caption{Box-whisker plot for the standard deviations obtained by the different models. Each model contributes its own standard deviation as a data point for the diagram. 
		Therefore, the leftmost column represents~$1,000$ data points (from~\cite{hu2018tetrahedral}), the center column represents~$100$ data points (from~\cite{choi2016large}), and the rightmost column represents~$100$ data points (from~\cite{bogo2014faust}).}
	\label{fig:BoxWhiskerSDK}
\end{figure*}

In order to interpret the neighborhood sizes, we consider a box-whisker plot over all standard deviations within the respective models in Figure~\ref{fig:BoxWhiskerSDK}.
That is to say, the boxes indicate the \rev{first, second (median), and third quartile} of the standard deviations of neighborhood sizes for the indicated model repository.
In particular, most approaches in the literature use---and are evaluated on---a setting with a fixed neighborhood size~$k$.
In our analysis, this would correspond to a standard deviation around~$0$, indicating no or small changes to the neighborhood size within a geometry.
However, it is obvious from Figure~\ref{fig:BoxWhiskerSDK} that all standard deviations are located well away from~$0$.
\rev{Even considering the minima, i.e., the lower whiskers of the boxes, they reside at $2.7$, $4.1$, and $4.4$, respectively, indicating that at least these small variations in neighborhood size are necessary to minimize the classification error.
Note that the variation of neighborhood size is most notable for the clean model repository, where the standard deviation of chosen neighborhood sizes goes up to $5.46$.
This is in contrast to the scanned data from~\cite{bogo2014faust}, where the models are not very diverse, which is reflected in the almost uniform standard deviation of the chosen neighborhood sizes.}

In summary, from the global and local analysis of the obtained neighborhood sizes~$k$, we draw the following conclusions:
\begin{itemize}
	\item All standard deviations lie well above~$0$, i.e., the considered classification error favors variable neighborhood sizes over constant-size neighborhoods.
	\item This behavior is more pronounced for scanned models (\cite{choi2016large} and \cite{bogo2014faust}) than for clean models (\cite{hu2018tetrahedral}).
	\item The classification error favors smaller neighborhood sizes for clean models, however for scanned models this behavior is not preserved.
\end{itemize}


\subsection{Application Scenario}
\label{sec:ApplicationScenario}

\noindent \rev{Building on the observation from the previous section that varying neighborhood sizes can contribute to better performance and in order to evaluate our proposed methodology in an application scenario, we turn to the normal filtering stage of the point set denoising algorithm proposed in~\cite{yadav2018constraint}. 
	This first stage is part of a larger, iterative process of three stages that removes noise from an input geometry.
	We focus on the first stage to not have the effect of our weighting scheme be confounded by procedures within the more complex pipeline (we provide the names of the parameters of the algorithms in brackets in the following).
	In each iteration (parameter $ p $), a weighted covariance matrix is built.
	The algorithm is using a sharp cut-off weight function (parameter $ \rho $), optimizes the eigenvalues of the covariance matrix (parameter $ \tau $), and uses those to update the respective point normals afterwards.}

\begin{figure*}
	\centering
	\begin{subfigure}[t]{0.13\textwidth}
		\includegraphics[width=1.\textwidth]{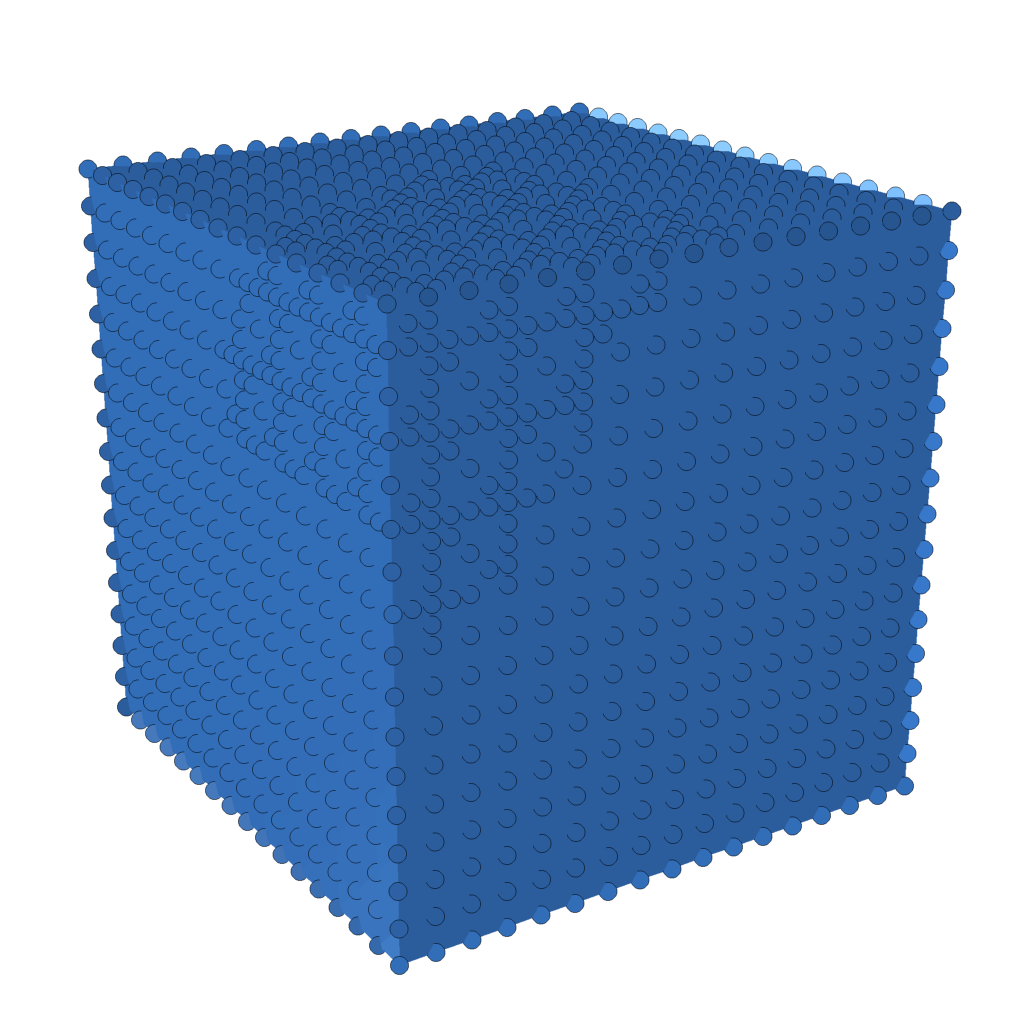}
		\caption{}
	\end{subfigure}
	\hfill
	\begin{subfigure}[t]{0.13\textwidth}
		\includegraphics[width=1.\textwidth]{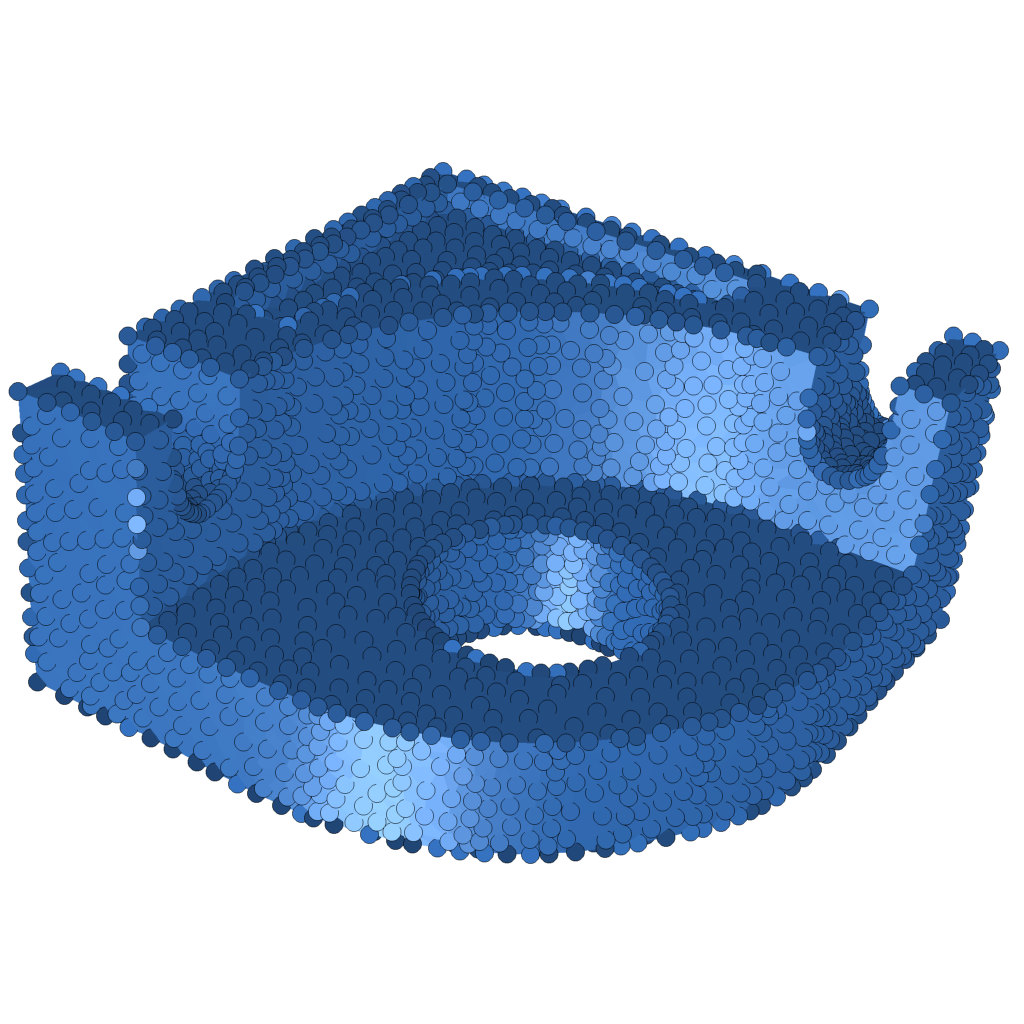}
		\caption{}
	\end{subfigure}
	\hfill
	\begin{subfigure}[t]{0.13\textwidth}
		\includegraphics[width=1.\textwidth]{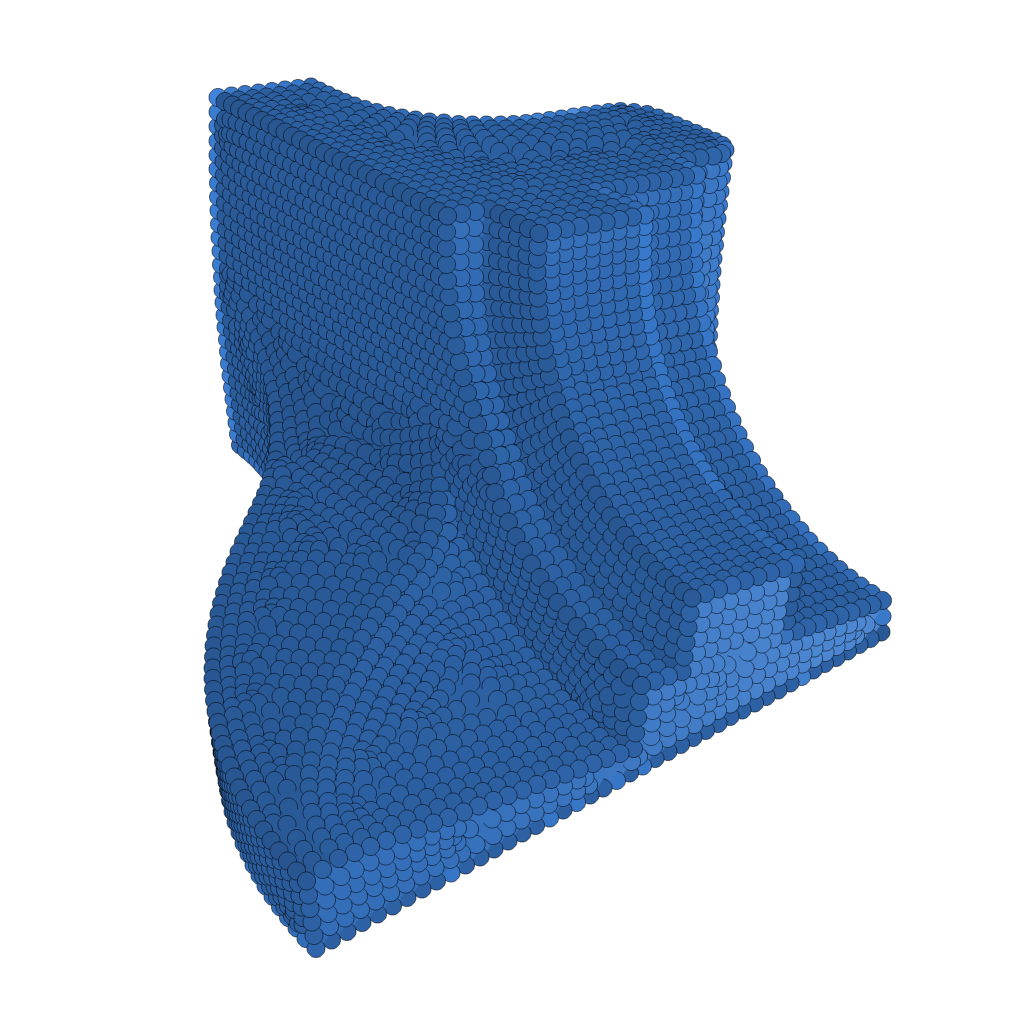}
		\caption{}
	\end{subfigure}
	\hfill
	\begin{subfigure}[t]{0.13\textwidth}
		\includegraphics[width=1.\textwidth]{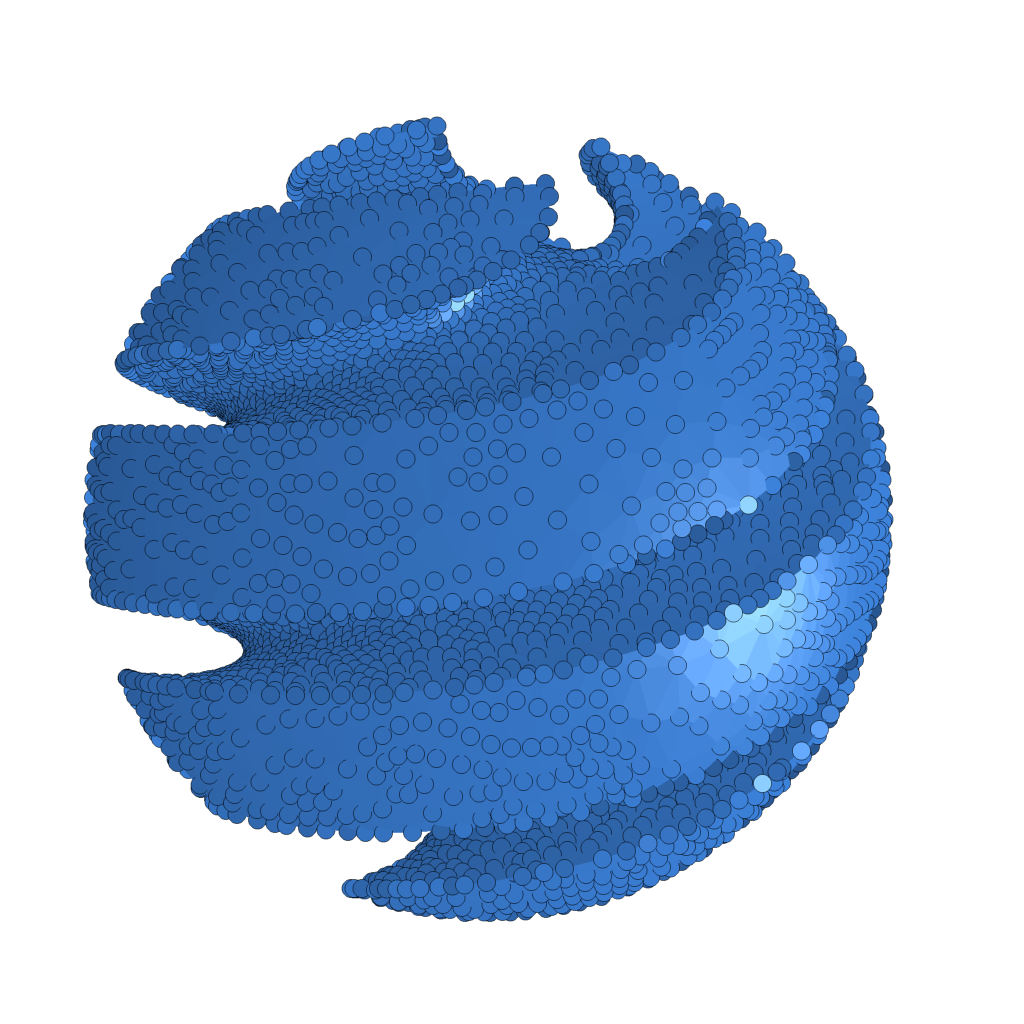}
		\caption{}
	\end{subfigure}
	\hfill
	\begin{subfigure}[t]{0.13\textwidth}
		\includegraphics[width=1.\textwidth]{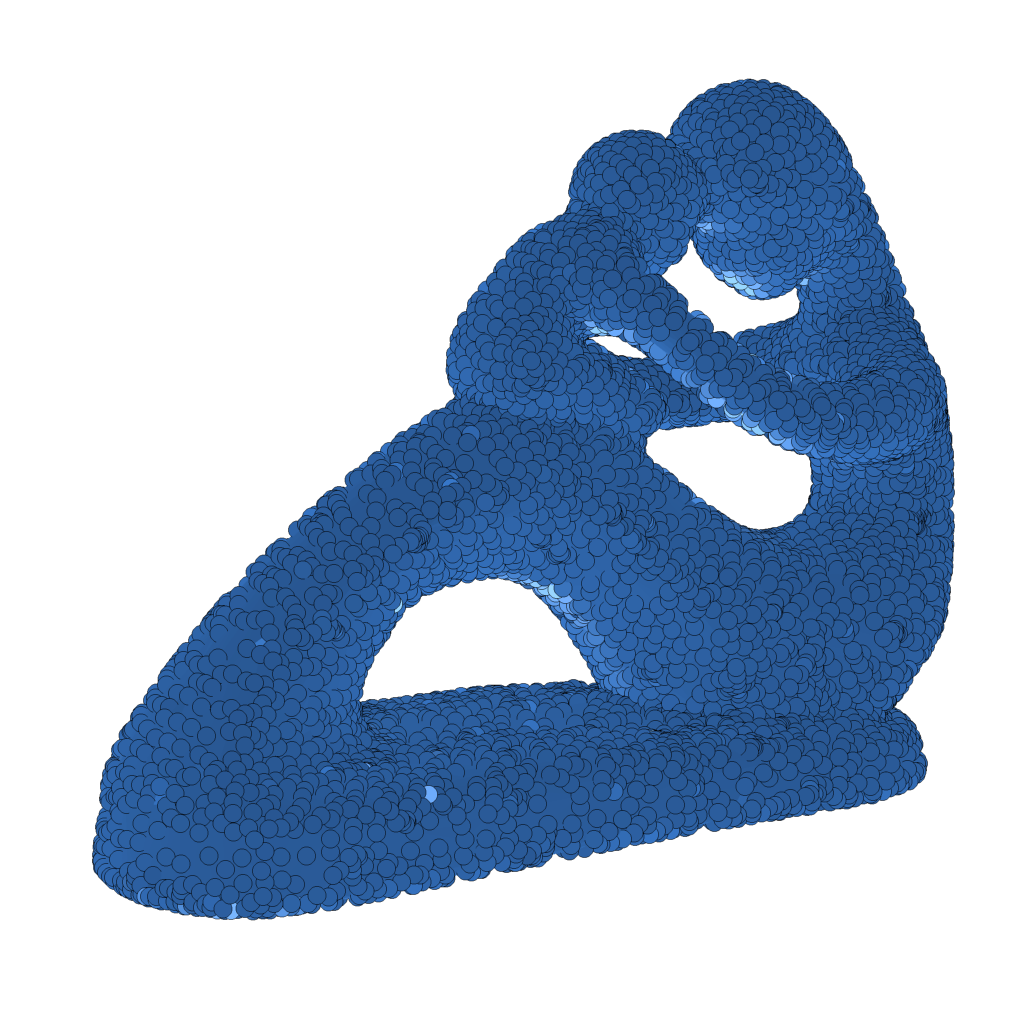}
		\caption{}
	\end{subfigure}
	\hfill
	\begin{subfigure}[t]{0.13\textwidth}
		\includegraphics[width=1.\textwidth]{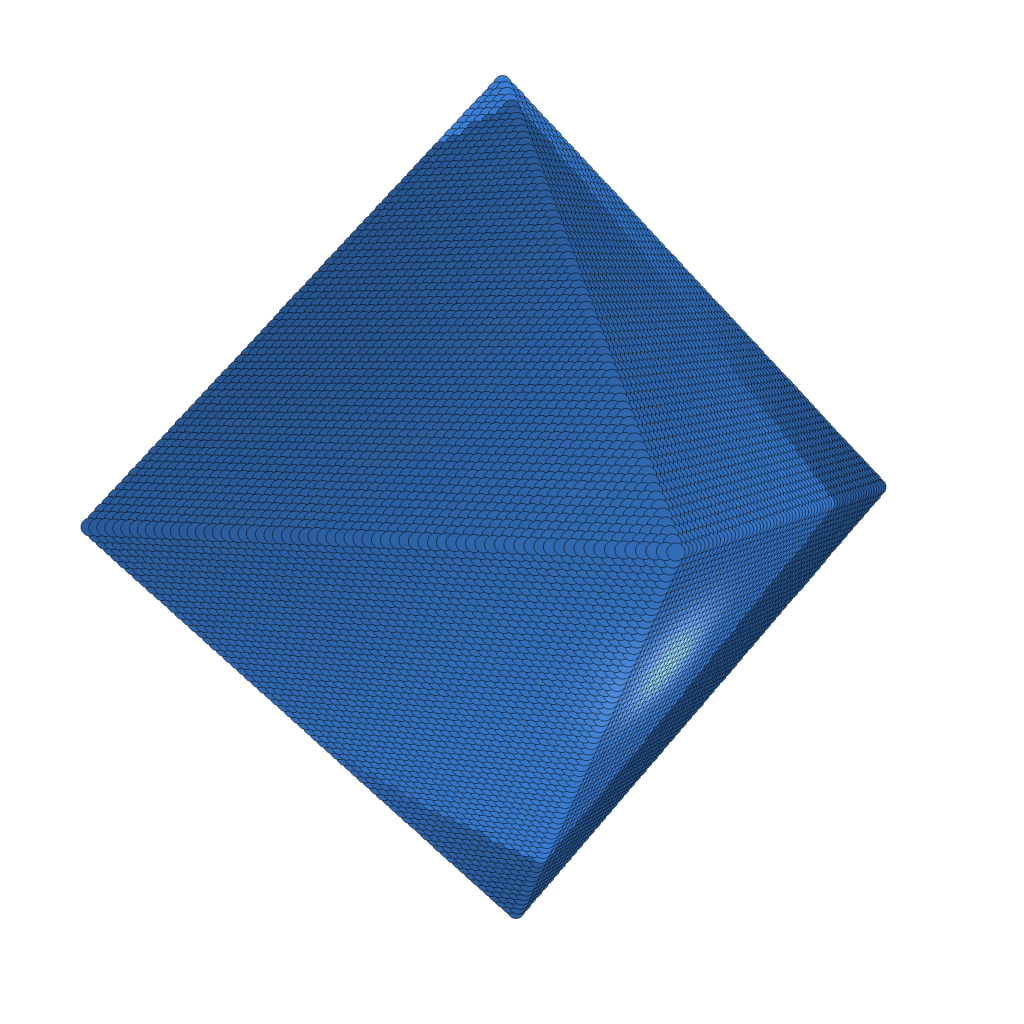}
		\caption{}
	\end{subfigure}
	\hfill
	\begin{subfigure}[t]{0.13\textwidth}
		\includegraphics[width=1.\textwidth]{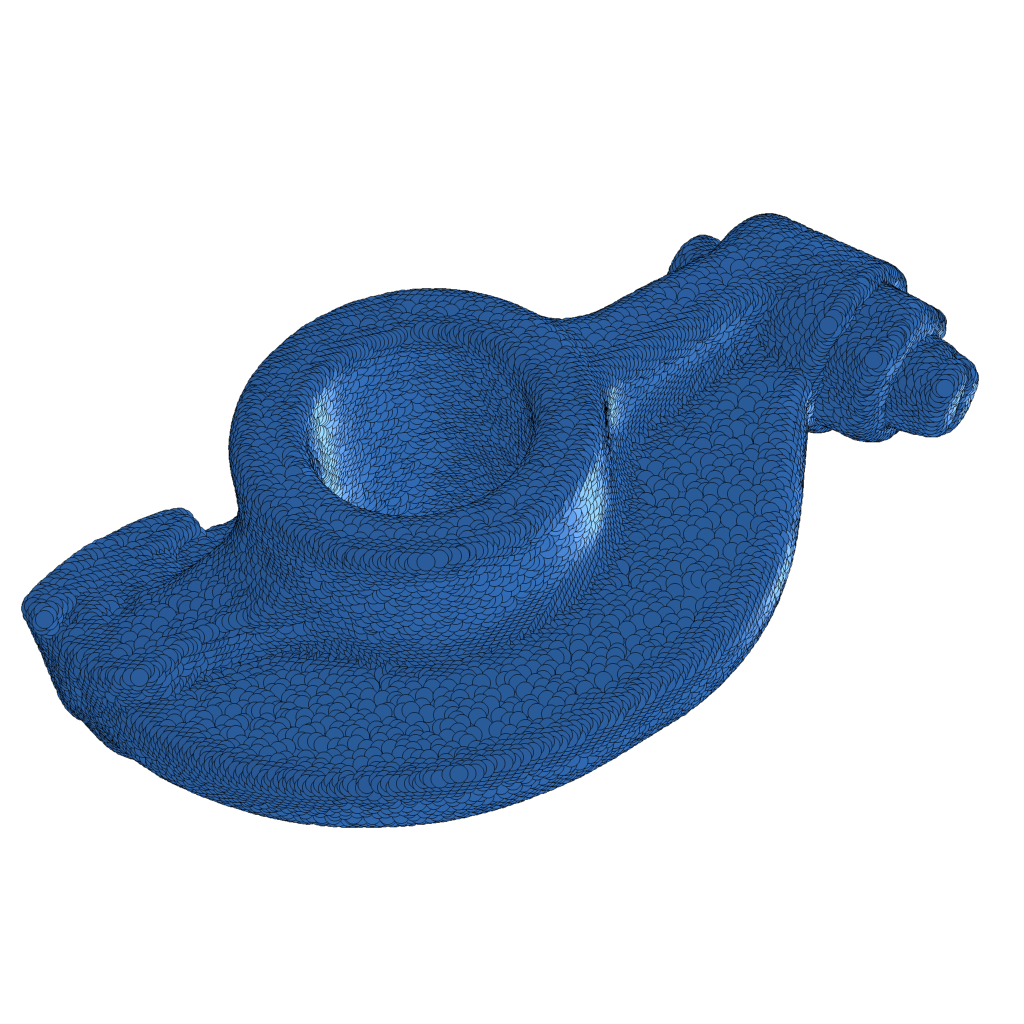}
		\caption{}
	\end{subfigure}
	\caption{Geometries from left to right with number of vertices in brackets: Cube ($ 1,906 $), Bearing ($ 3,475 $), Fandisk ($ 6,475 $), Sharp Sphere ($ 8,354 $), Fertility ($ 9,239 $), Octahedron ($ 16,395 $), and Rockerarm ($ 24,106 $).}
	\label{fig:applicationModels}
\end{figure*}

\rev{For the experiment, we followed the experimental pipeline of the original article~\cite{yadav2018constraint}.
Namely, we took the four geometries (Cube, Fandisk, Octahedron, Rockerarm) as discussed in the original publication~\cite{yadav2018constraint}, together with three more geometries (Bearing, Sharp Sphere, Fertility), see Figure~\ref{fig:applicationModels}.
The models were given as meshes and provided the oriented point normals, as they were obtained from the mesh information as weighted vertex normals that served as ground truth normals in the experiment. 
Afterwards, we applied Gaussian noise in normal direction with amplitudes~$ 0.3\ell $ and~$ 0.6\ell $, where~$ \ell $ denotes the mean of all distances from points to their respective six nearest neighbors. 
To measure the deviation from the ground truth normals, we make use of the mean squared error (MSE) given as}
\begin{align}
	\rev{\MSE(N, \tilde{N}) = \frac{1}{n} \sum_{i = 1}^{n} \lVert n_i - \tilde{n}_i \rVert^2_2,}
\end{align}
\rev{where $ N = \lbrace n_i \mid p_i\in P \rbrace $ are the ground truth and $ \tilde{N} = \lbrace \tilde{n}_i \mid p_i\in P \rbrace $ are the changed normals of the geometry~$P$.}

\rev{The default values for the normal filtering were taken from~\cite{yadav2018constraint} and the parameters $ (p, \tau, \rho) $ are set as follows:}
\begin{itemize}
	\item \rev{Cube with $ (150, 0.95, 0.3) $, }
	\item \rev{Fandisk with $ (150,0.9, 0.3) $, }
	\item \rev{Octahedron and Rockerarm with $ (80, 0.9, 025) $,}
	\item \rev{and Bearing, Sharp Sphere, and Fertility with $ (100, 0.9, 0.25) $.}
\end{itemize}
\rev{For the allowed neighborhood range, we first let the default normal filtering of~\cite{yadav2018constraint} identify the neighborhoods. 
Then, for each point~$p_i$, we took the number of neighbors~$k_i$ identified and let $ \mathfrak{K}_i:=\{k_i-10,\ldots,k_i+10\} $. 
If this caused negative values, we omitted them. 
Afterwards, each point chose a neighborhood according to the least error (Equation~(\ref{equ:Edim})) before and in each iteration. 
We set parameters $ a = b = \arccos(\rho) $ w.r.t.\@ the processed geometry, as this algorithm is tailored to benefit from a sharp distinction. 
In the cases where the covariance matrix degenerates as all weights become $ 0 $, we use the default neighborhood size and equal weights for such points.}

\rev{This allows us to study the effect of including our weighting scheme into a larger application by comparing to the results obtained by~\cite{yadav2018constraint}.
Furthermore, we can compare both the original results of~\cite{yadav2018constraint} and the results of our enhanced weighting pipeline to the ground truth normals provided by the noiseless models.
In Table~\ref{tab:mseResults}, the MSE results are shown, where MSE(Noise) compares ground truth and noisy input, MSE(\cite{yadav2018constraint}) compares ground truth and the result of~\cite{yadav2018constraint} with its default parameters, and MSE($ \mathfrak{K} $) compares ground truth and the result of the enhanced normal filtering pipeline.}

\begin{table*}
	\def\arraystretch{1.2}
	\begin{center}
		\begin{tabular}{l||ccc||ccc}
			& \multicolumn{3}{c||}{0.3$ \ell $} & \multicolumn{3}{c}{0.6$ \ell $} \\
			& MSE(Noise) & MSE(\cite{yadav2018constraint}) & MSE(ours) & MSE(Noise) & MSE(\cite{yadav2018constraint}) & MSE(ours)\\
			\hline\hline
			\bf Cube & 0.018171157 & 0.004720763 & \textbf{0.003817538} & 0.067785183 & 0.041014649 & \textbf{0.017185731}
			\\
			\hline
			\bf Bearing & \textbf{0.019149792} & 0.092202990 & 0.06311642&\textbf{0.087963467}&0.105885719&0.099716185
			\\
			\hline
			\bf Fandisk & \textbf{0.015985698} & 0.047105692 & 0.033043562 &0.059076477&0.067027229&\textbf{0.052858261}
			\\
			\hline
			\bf Sharp Sphere & \textbf{0.033954089} & 0.152198571 & 0.124993406 &\textbf{0.091947551}&0.162334741&0.143099697
			\\
			\hline
			\bf Fertility & \textbf{0.134043694} & 0.220368116 & 0.197418828&\textbf{0.232151395}&0.25364564&0.25100736
			\\
			\hline
			\bf Octahedron & 0.287851891 & 0.282346769 & \textbf{0.275334065} &0.304714745&0.277547626&\textbf{0.271626537}
			\\
			\hline
			\bf Rockerarm & \textbf{0.017150895} & 0.081370198 & 0.067150116&\textbf{0.063687905}&0.084070141&0.078171864
		\end{tabular}
	\end{center}
	\caption{Mean squared errors (MSE) for several models corrupted by noise $ 0.3\ell $ and $ 0.6\ell $ with $ \ell $ being the mean distance of all distances among six nearest neighbors. Our approach outperforms that of~\cite{yadav2018constraint} in all examples considered.}
	\label{tab:mseResults}
\end{table*}

From Table~\ref{tab:mseResults}, an immediate observation is that allowing individual neighborhood ranges reduces the overall error compared to the default values used in the normal filtering.
This becomes obvious as all values in the column $\MSE(\mathfrak{K})$ are smaller than those in column $\MSE$(\cite{yadav2018constraint}).
Thus, the algorithm benefits from individual neighborhood ranges as discussed in Section~\ref{sec:LocalKAnalysis}, instead of setting one global neighborhood size parameter.

\rev{One interesting observation can be made by incorporating the MSE(Noise) values into the comparison.
For most of the models considered, these values outperform the normal filtering results. 
To explain this, consider that the point normals are influenced by the mesh properties, i.e., the faces and their areas affect the normals provided as ground truth. 
The filtering procedures thus sometimes worsens the results as it tries to clearly arrange normals corresponding to points which represent a planar area, whereas points lying on (sharp) edges in a mesh receive normals neither belonging to the areas meeting at that edge.
This points towards future work and improvements of~\cite{yadav2018constraint}, but is irrelevant in the context of this paper as the main message to be taken from these experiments is that utilizing optimized neighborhood selection over a wider range of~$k$ does positively influence the performance of the algorithm.}


\section{Conclusion}
\label{sec:conclusion}

\noindent In this article, we investigated a family of weights (Eq.~(\ref{equ:Weights})) for point set processing. 
These weights are based on the normal similarity. 
The family includes common choices such as equal weights or sharp cut-off weights at a given threshold.
Furthermore, we presented an evaluation model for neighborhood weights based on a Shannon entropy classification error (Eq.~(\ref{equ:Edim})).
We have performed a large-scale evaluation of our weight family on four data sets. 
The first set consisted of~$1,000$ clean surface meshes from the work of~\cite{hu2018tetrahedral}. 
The second and third set consisted of~$100$ real-world scans taken from each~\cite{choi2016large} and \cite{bogo2014faust}. 
Additionally, we included a scanned model from \cite{laric2012three} with over one million points.

A statistical analysis revealed that the optimal weight parameters should lead to a neglect of non-similar normals, yet include mid-range normal points with a low weight.
Specifically, equal weights, as used in the literature discussed in Section~\ref{sec:RelatedWork} and in particular in~\cite{weinmann2014semantic} do not obtain minimal error values.
Furthermore, sharp cut-off weights as used, e.g., by~\cite{yadav2018constraint} do perform well on certain scanned models, but are also generally inferior to more flexible weighting terms.
Finally, it became obvious in the evaluation that neighborhood sizes have to be variable over a point set as only these variable sizes attain minimal error values. 
\rev{The potential of this variability of neighborhood sizes was shown by incorporating point-wise neighborhood size ranges within a normal filtering stage in a point denoising algorihtm (\cite{yadav2018constraint}), yielding smaller mean squared errors compared to the original pipeline.}

While this article addresses a variety of possible weighting choices and neighborhood sizes, to cover the most widely used versions from the literature, several aspects are left as future work.
Further research consists of running the large-scale analysis on a broader range of neighborhood sizes, comparable to~\cite{weinmann2014semantic}. 

\mpar{\label{rev:33}}\rev{Not only with the publication of the versatile \emph{PointNet} architecture, machine- and deep learning techniques gradually conquered the realm of point set processing~\cite{qi2017pointnet}.
Subsequently, a wide variety of publications arose that tackle several problems related to point set processing.
These touch on multiple areas discussed in this paper.
For instance, the \emph{PointCleanNet} architecture was designed for denoising and outlier removal of point sets~\cite{rakotosaona2020pointcleannet}.
Another representative of machine learning technology for point sets is the \emph{NormNet} architecture, which derives point-wise normal estimations for three-dimensional point sets~\cite{hyeon2019normnet}.}
\mpar{\label{rev:34}}\rev{A problem naturally approached using normal information on a point cloud is (semantic) segmentation.
In a sense, our optimized neighborhood weights do segment the point set into several, small parts with consistent normal information.
In contrast, large-scale segmentation approaches via machine learning are available, e.g., utilizing edge-convolution networks~\cite{contreras2019edge}.}
\rev{All these machine learning approaches have in common that they depend on large sets of well labeled training data.
In its generality, our approach can add to these developments in many ways.
Aside from improving the (semi-)automated generation of training data, it can support the selection of neighborhood weights and sizes that serve as architecture input.
Additionally, the direct incorporation of normal information on the neighborhood-level supports topological consistence, outlier resistance, and coherence of both the considered point set as well as the algorithmic procedure applied to it.
Investigating these potentials of our methodology in the machine learning context is, however, left for future work.}

\mpar{\label{rev:39}}\rev{As a closing remark, we would like to point out that several algorithms developed within the geometry processing community are heavily dependent on one or more parameters.
These parametric values are often fine-tuned during an experimentation phase that is run on a small and limited data set.
Our large-scale analysis in this paper highlights the importance of a systematic setup for parameter evaluation.
In particular, we were able to show that the usually globally chosen neighborhood size parameter yields better results when used within a bilateral weighting scheme that incorporates normal information and varying neighborhood sizes.
We hope that this work encourages further research on the applicability of algorithms and their parameters ``in the wild'' to ensure applicability and robustness of the developed methods.}

\newpage

\appendix

\printcredits

\bibliographystyle{cas-model2-names}

\bibliography{literature}

\begin{thebibliography}{34}
\expandafter\ifx\csname natexlab\endcsname\relax\def\natexlab#1{#1}\fi
\providecommand{\url}[1]{\texttt{#1}}
\providecommand{\href}[2]{#2}
\providecommand{\path}[1]{#1}
\providecommand{\DOIprefix}{doi:}
\providecommand{\ArXivprefix}{arXiv:}
\providecommand{\URLprefix}{URL: }
\providecommand{\Pubmedprefix}{pmid:}
\providecommand{\doi}[1]{\href{http://dx.doi.org/#1}{\path{#1}}}
\providecommand{\Pubmed}[1]{\href{pmid:#1}{\path{#1}}}
\providecommand{\bibinfo}[2]{#2}
\ifx\xfnm\relax \def\xfnm[#1]{\unskip,\space#1}\fi
\bibitem[{Alexa et~al.(2001)Alexa, Behr, Cohen-Or, Fleishman, Levin and
  Silva}]{alexa2001point}
\bibinfo{author}{Alexa, M.}, \bibinfo{author}{Behr, J.},
  \bibinfo{author}{Cohen-Or, D.}, \bibinfo{author}{Fleishman, S.},
  \bibinfo{author}{Levin, D.}, \bibinfo{author}{Silva, C.T.},
  \bibinfo{year}{2001}.
\newblock \bibinfo{title}{{Point Set Surfaces}}, in:
  \bibinfo{booktitle}{VIS'01: Proceedings of the conference on Visualization
  '01}, \bibinfo{organization}{IEEE Computer Society}. pp.
  \bibinfo{pages}{21--28}.
\bibitem[{Bellekens et~al.(2014)Bellekens, Spruyt, Berkvens and
  Weyn}]{bellekens2014survey}
\bibinfo{author}{Bellekens, B.}, \bibinfo{author}{Spruyt, V.},
  \bibinfo{author}{Berkvens, R.}, \bibinfo{author}{Weyn, M.},
  \bibinfo{year}{2014}.
\newblock \bibinfo{title}{A survey of rigid 3d pointcloud registration
  algorithms}, in: \bibinfo{booktitle}{AMBIENT 2014: the Fourth International
  Conference on Ambient Computing, Applications, Services and Technologies,
  August 24-28, 2014, Rome, Italy}, pp. \bibinfo{pages}{8--13}.
\bibitem[{Belton and Lichti(2006)}]{belton2006classification}
\bibinfo{author}{Belton, D.}, \bibinfo{author}{Lichti, D.D.},
  \bibinfo{year}{2006}.
\newblock \bibinfo{title}{Classification and segmentation of terrestrial laser
  scanner point clouds using local variance information}.
\newblock \bibinfo{journal}{The International Archives of the Photogrammetry,
  Remote Sensing, and Spatial Information Sciences} \bibinfo{volume}{36},
  \bibinfo{pages}{44--49}.
\bibitem[{Boehnen and Flynn(2005)}]{boehnen2005accuracy}
\bibinfo{author}{Boehnen, C.}, \bibinfo{author}{Flynn, P.},
  \bibinfo{year}{2005}.
\newblock \bibinfo{title}{Accuracy of {3D} scanning technologies in a face
  scanning scenario}, in: \bibinfo{booktitle}{IEEE Fifth International
  Conference on 3D Digital Imaging and Modeling.}, pp.
  \bibinfo{pages}{310--317}.
\bibitem[{Bogo et~al.(2014)Bogo, Romero, Loper and Black}]{bogo2014faust}
\bibinfo{author}{Bogo, F.}, \bibinfo{author}{Romero, J.},
  \bibinfo{author}{Loper, M.}, \bibinfo{author}{Black, M.J.},
  \bibinfo{year}{2014}.
\newblock \bibinfo{title}{{FAUST}: Dataset and evaluation for {3D} mesh
  registration}, in: \bibinfo{booktitle}{Proceedings IEEE Conf. on Computer
  Vision and Pattern Recognition (CVPR)}, \bibinfo{publisher}{IEEE},
  \bibinfo{address}{Piscataway, NJ, USA}. pp. \bibinfo{pages}{3794--3801}.
\bibitem[{Brodu and Lague(2012)}]{brodu2012terrestrial}
\bibinfo{author}{Brodu, N.}, \bibinfo{author}{Lague, D.}, \bibinfo{year}{2012}.
\newblock \bibinfo{title}{{3D} terrestrial lidar data classification of complex
  natural scenes using a multi-scale dimensionality criterion: Applications in
  geomorphology}.
\newblock \bibinfo{journal}{ISPRS Journal of Photogrammetry and Remote Sensing}
  \bibinfo{volume}{68}, \bibinfo{pages}{121--134}.
\bibitem[{Buck et~al.(2007)Buck, Naether, Braun, Bolliger, Friederich,
  Jackowski, Aghayev, Christe, Vock, Dirnhofer et~al.}]{buck2007application}
\bibinfo{author}{Buck, U.}, \bibinfo{author}{Naether, S.},
  \bibinfo{author}{Braun, M.}, \bibinfo{author}{Bolliger, S.},
  \bibinfo{author}{Friederich, H.}, \bibinfo{author}{Jackowski, C.},
  \bibinfo{author}{Aghayev, E.}, \bibinfo{author}{Christe, A.},
  \bibinfo{author}{Vock, P.}, \bibinfo{author}{Dirnhofer, R.}, et~al.,
  \bibinfo{year}{2007}.
\newblock \bibinfo{title}{Application of {3D} documentation and geometric
  reconstruction methods in traffic accident analysis with high resolution
  surface scanning, radiological {MSCT/MRI} scanning and real data based
  animation}.
\newblock \bibinfo{journal}{Forensic science international}
  \bibinfo{volume}{170}, \bibinfo{pages}{20--28}.
\bibitem[{Choi et~al.(2016)Choi, Zhou, Miller and Koltun}]{choi2016large}
\bibinfo{author}{Choi, S.}, \bibinfo{author}{Zhou, Q.Y.},
  \bibinfo{author}{Miller, S.}, \bibinfo{author}{Koltun, V.},
  \bibinfo{year}{2016}.
\newblock \bibinfo{title}{A large dataset of object scans}.
\newblock \bibinfo{journal}{arXiv:1602.02481} .
\bibitem[{Contreras and Denzler(2019)}]{contreras2019edge}
\bibinfo{author}{Contreras, J.}, \bibinfo{author}{Denzler, J.},
  \bibinfo{year}{2019}.
\newblock \bibinfo{title}{Edge-convolution point net for semantic segmentation
  of large-scale point clouds}, in: \bibinfo{booktitle}{IGARSS 2019 - 2019 IEEE
  International Geoscience and Remote Sensing Symposium}, pp.
  \bibinfo{pages}{5236--5239}.
\newblock \DOIprefix\doi{10.1109/IGARSS.2019.8899303}.
\bibitem[{Demantk{\'e} et~al.(2011)Demantk{\'e}, Mallet, David and
  Vallet}]{demantke2011dimensionality}
\bibinfo{author}{Demantk{\'e}, J.}, \bibinfo{author}{Mallet, C.},
  \bibinfo{author}{David, N.}, \bibinfo{author}{Vallet, B.},
  \bibinfo{year}{2011}.
\newblock \bibinfo{title}{Dimensionality based scale selection in {3D} lidar
  point clouds}.
\newblock \bibinfo{journal}{The International Archives of the Photogrammetry,
  Remote Sensing and Spatial Information Sciences}
  \bibinfo{volume}{XXXVIII-5/W12}, \bibinfo{pages}{97--102}.
\bibitem[{Floater and Reimers(2001)}]{floater2001meshless}
\bibinfo{author}{Floater, M.S.}, \bibinfo{author}{Reimers, M.},
  \bibinfo{year}{2001}.
\newblock \bibinfo{title}{Meshless parametrization and surface reconstruction}.
\newblock \bibinfo{journal}{Computer Aided Geometric Design}
  \bibinfo{volume}{18}, \bibinfo{pages}{77--92}.
\bibitem[{Hoppe et~al.(1992)Hoppe, DeRose, Duchamp, McDonald and
  Stuetzle}]{hoppe1992surface}
\bibinfo{author}{Hoppe, H.}, \bibinfo{author}{DeRose, T.},
  \bibinfo{author}{Duchamp, T.}, \bibinfo{author}{McDonald, J.},
  \bibinfo{author}{Stuetzle, W.}, \bibinfo{year}{1992}.
\newblock \bibinfo{title}{{Surface Reconstruction from Unorganized Points}},
  in: \bibinfo{booktitle}{Proceedings of the 19th annual conference on Computer
  graphics and interactive techniques}, \bibinfo{organization}{ACM}. pp.
  \bibinfo{pages}{71--78}.
\bibitem[{Hu et~al.(2018)Hu, Zhou, Gao, Jacobson, Zorin and
  Panozzo}]{hu2018tetrahedral}
\bibinfo{author}{Hu, Y.}, \bibinfo{author}{Zhou, Q.}, \bibinfo{author}{Gao,
  X.}, \bibinfo{author}{Jacobson, A.}, \bibinfo{author}{Zorin, D.},
  \bibinfo{author}{Panozzo, D.}, \bibinfo{year}{2018}.
\newblock \bibinfo{title}{Tetrahedral meshing in the wild.}
\newblock \bibinfo{journal}{ACM Trans. Graph.} \bibinfo{volume}{37},
  \bibinfo{pages}{60--1}.
\bibitem[{Hyeon et~al.(2019)Hyeon, Lee, Kim and Doh}]{hyeon2019normnet}
\bibinfo{author}{Hyeon, J.}, \bibinfo{author}{Lee, W.}, \bibinfo{author}{Kim,
  J.H.}, \bibinfo{author}{Doh, N.}, \bibinfo{year}{2019}.
\newblock \bibinfo{title}{Normnet: Point-wise normal estimation network for
  three-dimensional point cloud data}.
\newblock \bibinfo{journal}{International Journal of Advanced Robotic Systems}
  \bibinfo{volume}{16}, \bibinfo{pages}{1729881419857532}.
\bibitem[{Laric(2012)}]{laric2012three}
\bibinfo{author}{Laric, O.}, \bibinfo{year}{2012}.
\newblock \URLprefix \url{http://threedscans.com/}.
\bibitem[{Levin(1998)}]{levin1998approximation}
\bibinfo{author}{Levin, D.}, \bibinfo{year}{1998}.
\newblock \bibinfo{title}{The approximation power of moving least-squares}.
\newblock \bibinfo{journal}{Mathematics of Computation} \bibinfo{volume}{67},
  \bibinfo{pages}{1517--1531}.
\bibitem[{Levin(2004)}]{levin2004mesh}
\bibinfo{author}{Levin, D.}, \bibinfo{year}{2004}.
\newblock \bibinfo{title}{{Mesh-independent Surface Interpolation}}, in:
  \bibinfo{editor}{Brunnett, G.}, \bibinfo{editor}{Hamann, B.},
  \bibinfo{editor}{M{\"u}ller, H.}, \bibinfo{editor}{Linsen, L.} (Eds.),
  \bibinfo{booktitle}{Geometric modeling for scientific visualization}.
  \bibinfo{publisher}{Springer}, pp. \bibinfo{pages}{37--49}.
\bibitem[{Levoy et~al.(2000)Levoy, Pulli, Curless, Rusinkiewicz, Koller,
  Pereira, Ginzton, Anderson, Davis, Ginsberg et~al.}]{levoy2000digital}
\bibinfo{author}{Levoy, M.}, \bibinfo{author}{Pulli, K.},
  \bibinfo{author}{Curless, B.}, \bibinfo{author}{Rusinkiewicz, S.},
  \bibinfo{author}{Koller, D.}, \bibinfo{author}{Pereira, L.},
  \bibinfo{author}{Ginzton, M.}, \bibinfo{author}{Anderson, S.},
  \bibinfo{author}{Davis, J.}, \bibinfo{author}{Ginsberg, J.}, et~al.,
  \bibinfo{year}{2000}.
\newblock \bibinfo{title}{{The Digital Michelangelo Project: 3D Scanning of
  Large Statues}}, in: \bibinfo{booktitle}{Proceedings of the 27th annual
  conference on Computer graphics and interactive techniques}, pp.
  \bibinfo{pages}{131--144}.
\bibitem[{Levoy and Whitted(1985)}]{levoy1985use}
\bibinfo{author}{Levoy, M.}, \bibinfo{author}{Whitted, T.},
  \bibinfo{year}{1985}.
\newblock \bibinfo{title}{{The Use of Points as a Display Primitive}}.
\newblock \bibinfo{type}{Technical Report}. University of North Carolina.
\bibitem[{Linsen and Prautzsch(2001)}]{linsen2001local}
\bibinfo{author}{Linsen, L.}, \bibinfo{author}{Prautzsch, H.},
  \bibinfo{year}{2001}.
\newblock \bibinfo{title}{{Local Versus Global Triangulations}}, in:
  \bibinfo{booktitle}{Proceedings of EUROGRAPHICS}, pp.
  \bibinfo{pages}{257--263}.
\bibitem[{Lipman et~al.(2006)Lipman, Cohen-Or and Levin}]{lipman2006error}
\bibinfo{author}{Lipman, Y.}, \bibinfo{author}{Cohen-Or, D.},
  \bibinfo{author}{Levin, D.}, \bibinfo{year}{2006}.
\newblock \bibinfo{title}{{Error Bounds and Optimal Neighborhoods for MLS
  Approximation}}, in: \bibinfo{booktitle}{Proceedings of the fourth
  Eurographics symposium on Geometry processing},
  \bibinfo{organization}{Eurographics Association}. pp.
  \bibinfo{pages}{71--80}.
\bibitem[{Marler et~al.(2006)Marler, Gehrman, Martin and
  Ancoli-Israel}]{marler2006sigmoidally}
\bibinfo{author}{Marler, M.R.}, \bibinfo{author}{Gehrman, P.},
  \bibinfo{author}{Martin, J.L.}, \bibinfo{author}{Ancoli-Israel, S.},
  \bibinfo{year}{2006}.
\newblock \bibinfo{title}{The sigmoidally transformed cosine curve: a
  mathematical model for circadian rhythms with symmetric non-sinusoidal
  shapes}.
\newblock \bibinfo{journal}{Statistics in medicine} \bibinfo{volume}{25},
  \bibinfo{pages}{3893--3904}.
\bibitem[{Mitra et~al.(2004)Mitra, Nguyen and Guibas}]{mitra2004estimating}
\bibinfo{author}{Mitra, N.J.}, \bibinfo{author}{Nguyen, A.},
  \bibinfo{author}{Guibas, L.}, \bibinfo{year}{2004}.
\newblock \bibinfo{title}{{Estimating Surface Normals in Noisy Point Cloud
  Data}}.
\newblock \bibinfo{journal}{International Journal of Computational Geometry \&
  Applications} \bibinfo{volume}{14}, \bibinfo{pages}{261--276}.
\bibitem[{Park et~al.(2012)Park, Lee and Lee}]{park2012multi}
\bibinfo{author}{Park, M.K.}, \bibinfo{author}{Lee, S.J.},
  \bibinfo{author}{Lee, K.H.}, \bibinfo{year}{2012}.
\newblock \bibinfo{title}{Multi-scale tensor voting for feature extraction from
  unstructured point clouds}.
\newblock \bibinfo{journal}{Graphical Models} \bibinfo{volume}{74},
  \bibinfo{pages}{197--208}.
\bibitem[{Pauly et~al.(2002)Pauly, Gross and Kobbelt}]{pauly2002efficient}
\bibinfo{author}{Pauly, M.}, \bibinfo{author}{Gross, M.},
  \bibinfo{author}{Kobbelt, L.}, \bibinfo{year}{2002}.
\newblock \bibinfo{title}{{Efficient Simplification of Point-Sampled
  Surfaces}}, in: \bibinfo{booktitle}{Proceedings of the conference on
  Visualization'02}, \bibinfo{organization}{IEEE Computer Society}. pp.
  \bibinfo{pages}{163--170}.
\bibitem[{Pauly et~al.(2003)Pauly, Keiser, Kobbelt and Gross}]{pauly2003shape}
\bibinfo{author}{Pauly, M.}, \bibinfo{author}{Keiser, R.},
  \bibinfo{author}{Kobbelt, L.}, \bibinfo{author}{Gross, M.},
  \bibinfo{year}{2003}.
\newblock \bibinfo{title}{{Shape Modeling with Point-Sampled Geometry}}.
\newblock \bibinfo{journal}{ACM Transactions on Graphics} \bibinfo{volume}{22,
  3}, \bibinfo{pages}{641--650}.
\bibitem[{Qi et~al.(2017)Qi, Su, Mo and Guibas}]{qi2017pointnet}
\bibinfo{author}{Qi, C.R.}, \bibinfo{author}{Su, H.}, \bibinfo{author}{Mo, K.},
  \bibinfo{author}{Guibas, L.J.}, \bibinfo{year}{2017}.
\newblock \bibinfo{title}{Pointnet: Deep learning on point sets for 3d
  classification and segmentation}, in: \bibinfo{booktitle}{Proceedings of the
  IEEE conference on computer vision and pattern recognition}, pp.
  \bibinfo{pages}{652--660}.
\bibitem[{Rakotosaona et~al.(2020)Rakotosaona, La~Barbera, Guerrero, Mitra and
  Ovsjanikov}]{rakotosaona2020pointcleannet}
\bibinfo{author}{Rakotosaona, M.J.}, \bibinfo{author}{La~Barbera, V.},
  \bibinfo{author}{Guerrero, P.}, \bibinfo{author}{Mitra, N.J.},
  \bibinfo{author}{Ovsjanikov, M.}, \bibinfo{year}{2020}.
\newblock \bibinfo{title}{Pointcleannet: Learning to denoise and remove
  outliers from dense point clouds}, in: \bibinfo{booktitle}{Computer Graphics
  Forum}, \bibinfo{organization}{Wiley Online Library}. pp.
  \bibinfo{pages}{185--203}.
\bibitem[{Shannon(1948)}]{shannon1948mathematical}
\bibinfo{author}{Shannon, C.E.}, \bibinfo{year}{1948}.
\newblock \bibinfo{title}{{A Mathematical Theory of Communication}}.
\newblock \bibinfo{journal}{The Bell System Technical Journal}
  \bibinfo{volume}{27}, \bibinfo{pages}{379--423}.
\bibitem[{Skrodzki(2019)}]{skrodzki2019neighborhood}
\bibinfo{author}{Skrodzki, M.}, \bibinfo{year}{2019}.
\newblock \bibinfo{title}{Neighborhood Data Structures, Manifold Properties,
  and Processing of Point Set Surfaces}.
\newblock Ph.D. thesis. Freie Universit\"at Berlin. \bibinfo{address}{Berlin,
  Germany}.
\bibitem[{Skrodzki et~al.(2018)Skrodzki, Jansen and
  Polthier}]{skrodzki2018densityWeights}
\bibinfo{author}{Skrodzki, M.}, \bibinfo{author}{Jansen, J.},
  \bibinfo{author}{Polthier, K.}, \bibinfo{year}{2018}.
\newblock \bibinfo{title}{Directional density measure to intrinsically estimate
  and counteract non-uniformity in point clouds}.
\newblock \bibinfo{journal}{Computer Aided Geometric Design}
  \bibinfo{volume}{64}, \bibinfo{pages}{73 -- 89}.
\newblock \URLprefix
  \url{http://www.sciencedirect.com/science/article/pii/S0167839618300256},
  \DOIprefix\doi{https://doi.org/10.1016/j.cagd.2018.03.011}.
\bibitem[{Sober and Levin(2016)}]{sober2016manifolds}
\bibinfo{author}{Sober, B.}, \bibinfo{author}{Levin, D.}, \bibinfo{year}{2016}.
\newblock \bibinfo{title}{Manifolds' projective approximation using the moving
  least-squares {(MMLS)}}.
\newblock \bibinfo{journal}{CoRR} \bibinfo{volume}{abs/1606.07104}.
\newblock \URLprefix \url{http://arxiv.org/abs/1606.07104},
  \href{http://arxiv.org/abs/1606.07104}{\tt arXiv:1606.07104}.
\bibitem[{Weinmann et~al.(2014)Weinmann, Jutzi and
  Mallet}]{weinmann2014semantic}
\bibinfo{author}{Weinmann, M.}, \bibinfo{author}{Jutzi, B.},
  \bibinfo{author}{Mallet, C.}, \bibinfo{year}{2014}.
\newblock \bibinfo{title}{Semantic {3D} scene interpretation: a framework
  combining optimal neighborhood size selection with relevant features}.
\newblock \bibinfo{journal}{ISPRS Annals of the Photogrammetry, Remote Sensing
  and Spatial Information Sciences} \bibinfo{volume}{II-3},
  \bibinfo{pages}{181--188}.
\bibitem[{Yadav et~al.(2018)Yadav, Reitebuch, Skrodzki, Zimmermann and
  Polthier}]{yadav2018constraint}
\bibinfo{author}{Yadav, S.K.}, \bibinfo{author}{Reitebuch, U.},
  \bibinfo{author}{Skrodzki, M.}, \bibinfo{author}{Zimmermann, E.},
  \bibinfo{author}{Polthier, K.}, \bibinfo{year}{2018}.
\newblock \bibinfo{title}{Constraint-based point set denoising using normal
  voting tensor and restricted quadratic error metrics}.
\newblock \bibinfo{journal}{Computers \& Graphics} \bibinfo{volume}{74},
  \bibinfo{pages}{234--243}.
\newblock \URLprefix
  \url{http://www.sciencedirect.com/science/article/pii/S0097849318300797},
  \DOIprefix\doi{https://doi.org/10.1016/j.cag.2018.05.014}.

\end{thebibliography}

\vskip10em

\bio{img/Foto_MartinSkrodzki}
Martin Skrodzki has studied Mathematics and Computer Science at TU Dortmund University (Germany), Texas A\&M International University (USA), and Freie Universit\"at Berlin (Germany). He obtained his Dr.~rer.~nat. in 2019 at Freie Universit\"at Berlin under the supervision of Prof.~Polthier (Freie Universit\"at Berlin) and Prof.~Levin (Tel Aviv University, Israel). He has been a postdoctoral scholar at the Institute for Computational and Experimental Research in Mathematics (ICERM, Brown University, USA) and in the RIKEN Interdisciplinary Theoretical and Mathematical Sciences Program (RIKEN, Japan). He is currently a DFG-Walter-Benjamin postdoctoral fellow at the Computer Graphics and Visualization (CGV, TU Delft, the Netherlands).
\endbio

\bio{img/Foto_EricZimmermann}
Eric Zimmermann has studied Mathematics at Freie Universit\"at Berlin (Berlin, Germany). He currently is a doctoral candidate in the group ``Mathematical Geometry Processing'' under the guidance of Prof. Polthier (Freie Universit\"at Berlin, Germany), and he is part of the project C05 of the collaborative research cluster SFB Transregio 109 called ``Discretization in Geometry and Dynamics'' dealing with computational and structural aspects of point sets.
\endbio

\end{document}